\documentclass[aps,pra,preprint,10pt]{revtex4-2}
\usepackage[utf8]{inputenc}
\usepackage[T1]{fontenc}
\usepackage{graphicx}% Include figure files
\usepackage{dcolumn}% Align table columns on decimal point
\usepackage{fontawesome5}
\usepackage{amsmath,amsfonts,amsthm,bm}
\usepackage{wrapfig}
\usepackage{array}
\usepackage{soul}
\usepackage{etoolbox}
\usepackage{verbatim}
\usepackage{siunitx}
\usepackage[colorlinks=true]{hyperref}
\usepackage{booktabs}
\usepackage{bm}
\usepackage[version=4]{mhchem}
\usepackage{placeins} % allows for manual float barriers
\usepackage{dblfloatfix} % more reliable two-column tables
\newbool{Ldraft}
\booltrue{Ldraft}
\boolfalse{Ldraft}
\newbool{useNewHeader}
\booltrue{useNewHeader}

\begin{abstract}
Accurate prediction of the frequency response of quantum dots under incident electromagnetic radiation is essential for investigation of absorption spectra, excitonic effects, and nonlinear optical behavior in quantum dots and semiconductor nanoparticles. The polarization propagator provides a rigorous framework for evaluating these properties, but its construction is computationally demanding. Key challenges arise from the level of treatment of electron-electron correlation, the size of the excitonic basis, and the cost associated with evaluating two-electron integrals. These challenges intensify with increasing system size. In addition, explicit construction and storage of response matrices are often impractical because their increased dimensionality can lead to memory requirements that may exceed realistic computational limits. The present work addresses these challenges by developing both first-order and second-order frequency-dependent polarization propagator calculations for PbS and CdS quantum dots. The propagator is formulated using the electron propagator approach and is expressed as the resolvent of the Hamiltonian superoperator. Interaction between the nanoparticles and electromagnetic radiation was treated using the dipole approximation, and the light-matter interaction was expressed as a particle-hole excitation operator basis. The correlated ground state wavefunction was treated at MP2 level and all terms in the response matrix that were up to second-order in the fluctuating potential term were included in the calculation. A frequency-dependent inverse Krylov subspace method was derived and combined with the folded spectrum technique to isolate excitation energies within a chosen frequency window. This strategy avoids full diagonalization of the response matrix and significantly reduces computational cost for large systems. The method was implemented in a matrix-free fashion in which no explicit response matrix is assembled or stored, and all operations rely on matrix-vector products. This approach was applied to compute UV–VIS excitation spectra of PbS and CdS quantum dots. The results demonstrate that the frequency-dependent inverse Krylov-based subspace projection methods provide a computationally efficient yet accurate approximation for calculating excitation spectra of semiconductor quantum dots. The developed method enables correlated calculations for systems in which full diagonalization of the response matrix is computationally prohibitive.
\end{abstract}
\begin{document}
%\preprint{AIP/123-QED}
%\title{Calculation of Frequency-Dependent Polarization Propagator for Semiconductor Nanoparticles: Integrating Geminal-Screened Quasiparticle Kernel and Electron-Propagator Methods}
%\title{Calculation of Frequency-Dependent Polarization Propagator in Quantum Dots using Stratified Stochastic Diagrammatic Summation Method}
\title{Diagrammatic Evaluation of the Second Order Polarization Propagator Using Folded Spectrum}
\title{Frequency-Dependent Polarization Propagator Calculation for Quantum Dots Using Optimized Inverse Krylov Subspace and Folded-Spectrum Method}
\author{Chandler Martin}
\affiliation{Department of Physics, Syracuse University, Syracuse, New York 13244 USA}
\author{Nicole Spanedda}
\affiliation{Department of Chemistry, Syracuse University, Syracuse, New York 13244 USA}
\author{Anaira Jalan}
\affiliation{Department of Physics, Imperial College London, London SW7 2AZ, United Kingdom}
\author{Emily Schafer}
\affiliation{Department of Chemistry, Siena College, Loudonville, New York 12211 USA}
\author{Jessica Beyer}
\affiliation{Keck Science Department, Scripps College, Claremont, California 91711 USA}
\author{Arindam Chakraborty}%
\email{archakra@syr.edu}
\affiliation{Department of Chemistry, Syracuse University, Syracuse, New York 13244 USA}

\flushbottom
\maketitle
\thispagestyle{empty}
\section{Introduction}
The polarization propagator (PP) is a frequency-dependent response function that links the change in the electronic polarization of a system to the strength and frequency of an external perturbation. It provides a rigorous framework for describing excited-state properties and optical spectra, and it underpins many-body formulations of electron correlation. The poles of the PP correspond directly to excitation energies, while its residues determine transition strengths, making it a central object for the study of spectroscopic observables.
In the Lehmann representation, the PP is expressed as\cite{oddershede1978polarization}
\begin{align}
\label{eq:pp_def}
\Pi_{pq;rs}(\omega) = \lim_{\eta \rightarrow 0}\sum_{k}\frac{\langle \Psi_0 \vert p^\dagger q \vert \Psi_k \rangle
\langle \Psi_k \vert r^\dagger s \vert \Psi_0 \rangle}{\omega - (E_k - E_0) + i\eta},
\end{align}
where ${E_k,\Psi_k}$ are the exact eigenvalues and eigenstates of the electronic Hamiltonian.
Several approximate methods for computing the PP have been developed due to the pressing need for understanding of light-matter interaction properties in materials. The algebraic diagrammatic construction scheme (ADC) was originally introduced in 1983\cite{schirmer1983new}, and rewrites the PP in a non-diagonal basis as
\begin{align}
    \Pi(\omega) = \mathbf{F}^\dagger(\omega - \mathbf{M})^{-1}\mathbf{F}
\end{align}
then applies a perturbative expansion of the $\mathbf{f}$ and $\mathbf{M}$ matrices. This leads to solving the Hermitian eigenvalue problem
\begin{align}
    \mathbf{M}\mathbf{Y} = \mathbf{Y}\mathbf{\Omega}
\end{align}
which gives the excitation energies $\omega$ and transition amplitudes $\mathbf{y}$ as eigenvalues and eigenvectors of the $\mathbf{M}$ matrix\cite{dreuw2015algebraic,dreuw2023algebraic}. The ADC method has been used to calculate molecular response properties\cite{Trofimov2006algebraic}.
The second-order polarization propagator approximation (SOPPA) provides a computationally accessible formulation derived from an MP2 expansion of the polarization propagator
\begin{align}
    \langle \langle \mathbf{P;\mathbf{Q}}\rangle \rangle_\omega = (\mathbf{P}\vert \mathbf{h})(\mathbf{h}\vert \omega \hat{\mathbf{\mathbb{I}}} - \hat{\mathbf{\mathbb{H}}}\vert \mathbf{h})^{-1}(\mathbf{h}\vert \mathbf{Q})
\end{align}
which is expressed in the superoperator formalism originally presented for SOPPA\cite{oddershede1984polarization,Nielsen1980soppa}. This method has been used to calculate excitation spectra\cite{Packer1996soppa} and frequency dependent polarizabilities\cite{Jorgensen2020soppa}. 
\par
In addition to that the GW approximation is used to calculate the self energy in a many-body system
\begin{align}
    \Sigma \approx iGW.
\end{align}
This approximation is used in combination with the Bethe-Salpeter equations
% \begin{align}
%     \Gamma(P,p) = \int \frac{d^4k}{(2\pi)^4} K(P,p,k) S(k-\frac{p}{2}) \Gamma(P,k)S(k+\frac{p}{2})
% \end{align}
to give the combined approximation GW-BSE, which has become the favored tool for calculation of quasiparticle excitations in solids\cite{Ulpe2020gwbse,Hung2016gwbse}. 
The electron propagator is defined by
\begin{align}
    G(x,x';t-t') = i\Theta(t'-t)\langle N\vert \psi^\dagger(x',t')\psi(x,t)\vert N \rangle, \quad t' > t
\end{align}
and is a Green's function based method that is widely used for calculation of ionization energies and electron affinities\cite{Trofimovelecpropion}. The propagator also contains information about excitation energies and has been used to calculate optical spectra\cite{Schneider2015elecpropagator,Purvis1974elecpropagator,Ortiz2017elecpropagator}. 
Finally, equation-of-motion coupled-cluster is a computationally expensive, but highly accurate method for calculation of excited state phenomena. It is defined by the equations\cite{Stanton1993eomcc}
\begin{align}
    \vert \Psi_{\mathrm{exc}}\rangle &= \mathcal{R}\vert \Psi_0\rangle \\
    \vert \Psi_0 \rangle &= e^T\vert \Phi_0 \rangle
\end{align}
and have shown applications to calculation of absorption spectra\cite{Korona2003eomccsd}. 
\par
All of these methods have been great advancements in the modern understanding of excited states, but they remain hampered by scalability issues. In particular, as system size grows many of these methods require full construction and diagonalization of the many-body Hamiltonian, which is computationally prohibitive.
\begin{table}[h]
\begin{ruledtabular}
\caption{Size of A Matrix for different chemical systems}
\label{tab:amat_size}
\begin{tabular}{c c c c c c c} 
 \rule[-1ex]{0pt}{2.5ex} Chemical System & & 1p1h & 2p2h & 3p3h \\
  & Diameter (nm)  & $N_{\mathrm{occ}}\times N_{\mathrm{vir}}$ & $N_{\mathrm{occ}}^2\times N_{\mathrm{vir}}^2$ & $N_{\mathrm{occ}}^3\times N_{\mathrm{vir}}^3$ \\
 \hline
 \ce{Pb4S4} & 0.16 & $8.8\times 10^2$ & $7.74\times 10^5$ & $6.81\times 10^8$ \\
  
 \ce{Pb44S44} & 0.79 & $1.06\times 10^5$ & $1.13\times 10^{10}$ & $1.21\times 10^{15}$\\
 
 \ce{Pb140S140} & 2.08 & $9.1\times 10^5$ & $8.28\times 10^{11}$ & $7.53 \times 10^{17}$ \\
  
 \ce{Cd24S24} & 1.06 & $9.84\times 10^4$ & $9.7\times 10^9$ & $9.56\times 10^{16}$ \\
 
 \ce{Cd45S45} & 1.29 & $3.46\times 10^5$ & $1.19\times 10^{11}$ & $4.15\times 10^{16}$ \\
 
\end{tabular}
\end{ruledtabular}
\end{table}

Much of this computational challenge comes down to the expensive calculation of eigenvalues from large matrices. As shown in \autoref{tab:amat_size}, as system size increases the number of particle-hole states used in the excitation matrices increases rapidly. While the full Hamiltonian contains information about everything in the system, in many cases information is only needed about eigenvalues within a certain range, for example, to predict the energy properties of a material in the UV-VIS range it is only necessary to calculate eigenvalues of the Hamiltonian in that range. Full diagonalization involves spending a significant amount more computational effort for information that does not affect the intended property. This is referred to as the interior eigenvalue problem, and a variety of strategies have been explored. Filter diagonalization techniques have already been applied to physical systems\cite{Yu1999filterdiag}, and quantum mechanical models\cite{Wall1995filterdiag,Pieper2016filterdiag}. The folded spectrum method reformulates the eigenvalue problem in terms of squared shifts that emphasize eigenvalues near a desired value\cite{Wang1994foldedspec,Canning2000foldedspec}. While these methods are efficient, they often suffer from numerical instabilities or high computational overhead due to preconditioning.
\par
In this work, we introduce an alternative route for representing  the polarization propagator in terms of a resolvent superoperator, evaluated using an inverted Krylov subspace for construction of the frequency dependent basis. This basis is then applied using the folded spectrum method, a robust approach to the interior eigenvalue problem because the resolvent superoperator generating the subspace projection isolates eigenvalues near chosen values. This method naturally filters out the unwanted parts of the spectrum and focuses the computation on the eigenvalues of interest by creating the optimal basis for calculation of a given $\lambda$ near a given $\omega$. This allows for major gains in efficiency. This method is closely related to an analytical approach, the Riesz projector method, which has already been shown in light-matter interaction properties\cite{Zschiedrich2018rieszproj} and the connection between the two methods is described in \autoref{sec:riesz_proj}. The superoperator definition of the polarization propagator is detailed in \autoref{sec:pp_def}, and the excitation energies are calculated using the folded spectrum method in the inverse Krylov basis, which is described in \autoref{sec:krylov_theory}. The matrix elements are then evaluated using diagrammatics to discover non-contributing terms in the higher order corrections of the M-operator. The resulting two-electron integrals are evaluated using Monte-Carlo integration and compared with the $0^{\mathrm{th}}$ order Hartree-Fock (HF) absorption spectra. This method is tested on a range of CdS and PbS semiconductor quantum dots, with diameters ranging from $0.1 - 2$ nanometers. Our formulation uses the analytically derived expressions for higher order correlated effects in the polarization propagator in conjunction with efficient approximations like the folded spectrum to extend PP calculations to larger and more complex quantum systems.

\section{Theoretical and Computational Method}
\label{sec:theory}
\subsection{Non-diagonal representation of the polarization propagator in correlated basis}
\label{sec:pp_def}
The time-independent two-particle Green's function is defined as,
\begin{align}
\label{eq:full_pp_resolvent}
    \Pi_{pq;rs}(\omega)
    &=
    \langle \Psi_0 \vert p^\dagger q 
    \frac{1}{\omega - (H - E_0) + i\eta
    }r^\dagger s \vert \Psi_0 \rangle
\end{align}
where, $\Psi_0$ is the exact ground state of the Hamiltonian $H$ and $E_0$ is the corresponding ground state energy. 
The exact and approximate eigenspectum is defined as:
\begin{align}
    H_0 \Phi_n &= E_n^0 \Phi_n \\
    H \Psi_n &= E_n \Psi_n
\end{align}
The full two-particle propagator in Lehmann representation is defined as:
\begin{align}
\label{eq:full_pp}
    \Pi_{pq;rs}(\omega)
    &=
    \sum_{k} 
    \frac{
        \langle \Psi_0 \vert p^\dagger q \vert \Psi_k \rangle
        \langle \Psi_k \vert r^\dagger s \vert \Psi_0 \rangle
    }{
    \omega - (E_k - E_0) + i\eta
    }.
\end{align}
% The zeroth-order version is given as
% \begin{align}
%     \Pi_{pq;rs}^0(\omega)
%     &=
%     \sum_{k} 
%     \frac{
%         \langle \Phi_0 \vert p^\dagger q \vert \Phi_k \rangle
%         \langle \Phi_k \vert r^\dagger s \vert \Phi_0 \rangle
%     }{
%     \omega - (E_k^0 - E_0^0) + i\eta
%     }
% \end{align}
The Lehmann representation uses the energy eigenstates as the underlying basis and in this basis, matrix representation of the polarization propagator takes a very simple form which is given as
\begin{align}
    \bm{\Pi}(\omega) = \bm{X}^\dagger (\omega - \bm{\Omega})^{-1} \bm{X}, 
\end{align}
where
\begin{align}
    X_{k} = \langle \Psi_k \vert r^\dagger s \vert \Psi_0\rangle  
\end{align}
and
\begin{align}
    \Omega_{kk'} = (E_k-E_0) \delta_{kk'} .
\end{align}
Although, the Lehmann representation is very common and convenient it is not the only representation. In the ADC formulation, Schirmer et al. introduced the non-diagonal representation\cite{schirmer1983new} where the polarization propagator is represented in the non-diagonal basis,
\begin{align}
\label{eq:pp_adc}
    \bm{\Pi}(\omega) = \bm{F}^\dagger (\omega - \bm{M})^{-1} \bm{F} 
\end{align}
where,
\begin{align}
\label{eq:mkk_def}
    M_{k'k} = \langle \tilde{\Psi}_{k'} \vert H - E_0 \vert \tilde{\Psi}_k \rangle
\end{align}
and
\begin{align}
    F_k = \langle \tilde{\Psi}_k \vert r^\dagger s \vert \Psi_0\rangle.
\end{align}
Here, the set $\tilde{\Psi}_k$ is specifically chosen not to be the eigenvectors of $H$,
\begin{align}
    H \tilde{\Psi}_k \neq \lambda_k \tilde{\Psi}_k. 
\end{align}
There are various ways to construct the set of basis functions and previous work in ADC and electron-propagator method and EOM-CC have demonstrated the effectiveness of generating the basis by performing particle-hole excitation and de-excitation from a correlated ground state wavefunction. Specifically, it has been shown by Simons and J\o{}rgensen\cite{pouljoergensen2012} that,
\begin{align}
\label{eq:tk_def}
    \vert \tilde{\Psi}_k \rangle = T^\dagger_k \vert \Psi_0\rangle
\end{align}
where
\begin{align}
    T_k^\dagger = [\{a^\dagger i\},\{a^\dagger i\}^\dagger,\{a^\dagger b^\dagger j i\},\{a^\dagger b^\dagger j i\}^\dagger,\dots,]
\end{align}
will form a complete basis.
Substituting \autoref{eq:tk_def} into \autoref{eq:mkk_def} we get
\begin{align}
\label{eq:mkk_2}
    M_{k'k} 
    =
    \langle \Psi_0 \vert T_{k'} (H - E_0) T_k^\dagger \vert \Psi_0 \rangle.
\end{align}
We note that the expression $(H - E_0) T_k^\dagger \vert \Psi_0 \rangle$ can be
written in terms of the following commutation,
\begin{align}
\label{eq:comm}
    (H - E_0) T_k^\dagger \vert \Psi_0 \rangle
    &= 
        H T_k^\dagger \vert \Psi_0 \rangle - E_0 T_k^\dagger \vert \Psi_0 \rangle \\
    &=
        H T_k^\dagger \vert \Psi_0 \rangle - T_k^\dagger H \vert \Psi_0 \rangle \\
    &=
        [H,T_k^\dagger] \vert \Psi_0 \rangle.
\end{align}
Substituting \autoref{eq:comm} in \autoref{eq:mkk_2}, we get the superoperator representation of the matrix element of $\bm{M}$
\begin{align}
\label{eq:mkk_so}
    M_{k'k} = \langle \Psi_0 \vert T_{k'} [H,T_k^\dagger] \vert \Psi_0\rangle.    
\end{align}
The commutator $[H,T_k^\dagger]$ in \autoref{eq:mkk_so} is known as the superoperator in electron-propagrator theory\cite{Ortiz2013EPTreview} and as an adjoint operator in Lie algebra\cite{Gilmore2006Lie}
\begin{align}
   [H,T_k^\dagger] \equiv \widehat{H} T_k^\dagger \equiv 
    ad_H T_k^\dagger \equiv
    \mathcal{L}_H T_k^\dagger .
\end{align}
It also appears in quantum dynamics in the Lindbald equation of motion\cite{BreuerPetruccione2007}.
Practical implementation of \autoref{eq:pp_adc} requires approximation of the correlated ground states $\Psi_0$ and truncation of the operator space $T_k$ which are discussed in the following section.
\subsection{Second-order approximation of the polarization propagator}
\label{sec:second_order}
% The M-operators
% \begin{align}
% \label{eq:m_operator}
%     \mathbf{M}_{22}(\omega) 
%     =
%     \begin{bmatrix}
%      \mathbf{A} + \omega \mathbf{S}   & \mathbf{B} \\
%      \mathbf{B}                & \mathbf{A} - \omega \mathbf{S}
%     \end{bmatrix}
% \end{align}
% The $Q^\dagger$ is the set of all particle-hole creation operator,
% \begin{align}
%     Q^\dagger = \{ a^\dagger i\}
% \end{align}
% \begin{align}
%     (P|Q) = \langle 0 \vert [P^\dagger,Q] \vert 0 \rangle
% \end{align}
% Using this,
In this work, we use the second-order M\o{}ller--Plessett perturbation theory (MP2) to include electron-electron correlation and obtain the correlated ground state wavefunction $\Psi_0$.
Starting with the Hartree-Fock equation, the eigensolution of the Fock operators is used to obtain the single-particle states
\begin{align}
    f \vert p \rangle = \epsilon_p \vert p \rangle,
\end{align}
\begin{align}
    H_0 = \sum_{p} \epsilon_p p^\dagger p,
\end{align}
and the non-interacting ground state is defined as $\vert 0 \rangle$. The normal-ordered interacting Hamiltonian is defined as 
\begin{align}
    H_N \equiv H - \langle 0 \vert H \vert 0 \rangle
\end{align}
\begin{align}
    H_N = H_0 + W_N
\end{align}
where $W_N$ is the normal-ordered fluctuating potential and is defined as
\begin{align}
    W_N 
    = 
    \frac{1}{4} \sum_{pqrs} \langle pq \vert r_{12}^{-1} \vert rs \rangle_A 
    \{p^\dagger q^\dagger s r\}.
\end{align}
The ground state MP2 wavefunction is defined as
\begin{align}
    \vert \Psi_0 \rangle = \vert 0 \rangle + \vert \Psi^{(1)} \rangle
\end{align}
where the first-order correction is defined with respect to the zeroth-order resolvent $R_0$
\begin{align}
    \vert \Psi^{(1)} \rangle = R_0 W_N \vert 0 \rangle
\end{align}
\begin{align}
    R_0 = \sum_{p \neq 0} \vert \Phi_p \rangle \frac{1}{E_0^{(0)}-H_0} \langle \Phi_p \vert.
\end{align}
\par
In this work, we are interested in dipole-allowed particle hole excitation and restrict the $T_k^\dagger$ space to only 1-particle 1-hole (1p-1h) excitation and de-excitation operators,
\begin{align}
    T_k^\dagger = \{\{a^\dagger i\},\{i^\dagger a\}\}. 
\end{align}
The superoperator representation derived in \autoref{eq:mkk_so} allows
simplified evaluation of the matrix elements which is derived as follows.
We start by noting that,
\begin{align}
\label{eq:h_tk_commutator}
    [H,T_k^\dagger] = [H_N,T_k^\dagger].
\end{align}
The commutation of the one-body operator in $H_N$ can be evaluated as
\begin{align}
\label{eq:one_body}
    [p^\dagger q,x^\dagger y]
    &= \delta_{qx}p^\dagger y - \delta_{py}x^\dagger q.
\end{align}
Similarly, the two-body operator can be evaluated as
\begin{align}
\label{eq:two_body}
    [p^\dagger q^\dagger s r,x^\dagger y] 
    = 
      \delta_{rx}p^\dagger q^\dagger s y 
    - \delta_{sy} p^\dagger q^\dagger x^\dagger r  
    + 
      \delta_{qx}p^\dagger y s r 
    - \delta_{py} x^\dagger q^\dagger s r .
\end{align}
The $\bm{M}$ matrix can be block-factored in excitation and de-excitation blocks
\begin{align}
    \mathbf{M} 
    =
    \begin{bmatrix}
     \mathbf{A} & \mathbf{B} \\
     \mathbf{B} &  \mathbf{A}
    \end{bmatrix}
\end{align}
The eigenvalues for the $\bm{M}$ matrix can be obtained from the solution of
the following generalized eigenvalue equation
\begin{align}
\label{eq:m_eigen}
    \begin{bmatrix}
     \mathbf{A} & \mathbf{B} \\
     \mathbf{B} &  \mathbf{A}
    \end{bmatrix}
    \begin{bmatrix}
        \mathbf{X} \\
        \mathbf{Y}
    \end{bmatrix}
    = 
    \omega
    \begin{bmatrix}
     \mathbf{S} & 0 \\
     0 &  -\mathbf{S}
    \end{bmatrix}
    \begin{bmatrix}
        \mathbf{X} \\
        \mathbf{Y}
    \end{bmatrix}.
\end{align}
Because all the basis functions are constructed from 1p-1h excitation and de-excitations from the MP2 wavefunction, all the matrix elements in \autoref{eq:m_eigen} can be expressed as a sum of zeroth, first, and second order correction terms,
\begin{align}
    S_{ia,jb} = S_{ia,jb}^{(0)} + S_{ia,jb}^{(1)} + S_{ia,jb}^{(2)} 
\end{align}
\begin{align}
    A_{ia,jb} = A_{ia,jb}^{(0)} + A_{ia,jb}^{(1)} + A_{ia,jb}^{(2)} 
\end{align}
\begin{align}
    B_{ia,jb} = B_{ia,jb}^{(0)} + B_{ia,jb}^{(1)} + B_{ia,jb}^{(2)} 
\end{align}
The zeroth-order overlap is a direct consequence of the orthogonality of eigenstates of $H_0$ and is given as
\begin{align}
    S_{ia,jb}^{(0)} = \delta_{ij}\delta_{ab}.
\end{align}
The first-order to the overlap is also zero because of the orthogonality of the first-order correction with the Fermi-vacuum
\begin{align}
    S_{ia,jb}^{(1)} = 0.
\end{align}
The second-order correction to the overlap is given by
\begin{align}
\label{eq:s(2)}
    S_{ia,jb}^{(2)}
    &=
    \langle 0 \vert W_N R_0  \{i^\dagger a\} \{b^\dagger j\} R_0 W_N \vert 0\rangle
\end{align}
where $\{\dots\}$ represents normal ordering of the second quantized operators in the brackets.
% \begin{align}
%     \vert \Psi^{(1)} \rangle = R_0 W_N \vert 0 \rangle
% \end{align}
% \begin{align}
%     S_{ia,jb}^{(2)}
%     &=
%     \langle \Psi^{(1)} \vert \{i^\dagger a\} \{b^\dagger j\} \vert \Psi^{(1)}\rangle \\\notag
%     & \quad - \langle \Psi^{(1)} \vert \{b^\dagger j\} \{i^\dagger a\}  \vert \Psi^{(1)}\rangle \\
%     &=
%     \langle 0 \vert W_N R_0  \{i^\dagger a\} \{b^\dagger j\} R_0 W_N \vert 0\rangle \\\notag
%     & \quad- \langle 0 \vert W_N R_0  \{i^\dagger a\} \{b^\dagger j\} R_0 W_N \vert 0\rangle \\
%     &=
%     [W_N R_0  \{i^\dagger a\} \{b^\dagger j\} R_0 W_N]_\mathrm{linked} \\\notag
%     & \quad-[W_N R_0  \{b^\dagger j\} \{i^\dagger a\} R_0 W_N]_\mathrm{linked}
% \end{align}
% Which are given as,
% \begin{align}
%     S_{ia,jb}^{(2)} 
%     = 
%     \frac{1}{2} \delta_{ab}
%     \sum_{cdk} K_{ik}^{cd}K_{jk}^{cd}
%     -\frac{1}{2} \delta_{ij}
%     \sum_{ckl} K_{kl}^{ac} K_{kl}^{bc}
% \end{align}
The zeroth and first-order correction to the $\bm{A}$ matrix is given as
\begin{align}
\label{eq:a(0)}
    A_{ia,jb}^{(0)} = (\epsilon_a - \epsilon_i) \delta_{ij} \delta_{ab},
\end{align}
\begin{align}
\label{eq:a(1)}
    A_{ia,jb}^{(1)} = \langle i a \vert r_{12}^{-1} \vert b j \rangle_A, 
\end{align}
where the subscript $\langle\dots\rangle_A$ represent anti-symmetrized matrix elements.
The expressions \autoref{eq:a(0)} and \autoref{eq:a(1)} matches the conventional unscreened
electron-hole interaction found in Configuration Interaction Singles (CIS) time-dependent Hartree-Fock (TDHF) formulation. The second-order contribution to the $\bm{A}$ matrix is given as
\begin{align}
\label{eq:a(2)}
    A_{ia,jb}^{(2)}
    &=
    \langle 0 \vert W_N R_0 
    [\{i^\dagger a\},[H_0,\{b^\dagger j\}]] 
    R_0 W_N\vert 0\rangle .  
\end{align}
% \begin{align} \label{amat_eq}
%     A_{ia,jb}^{(2)} 
%     &=
%     \langle \Psi^{(1)} \vert 
%     [\{i^\dagger a\},[H_0,\{b^\dagger j\}]] 
%     \vert \Psi^{(1)}\rangle \\
%     &=
%     \langle 0 \vert W_N R_0 
%     [\{i^\dagger a\},[H_0,\{b^\dagger j\}]] 
%     R_0 W_N\vert 0\rangle \\
%     &=
%     [W_N R_0 
%     [\{i^\dagger a\},[H_0,\{b^\dagger j\}]] 
%     R_0 W_N]_\mathrm{linked}     
% \end{align}
The zeroth and first-order contributions to the $\bm{B}$ matrix are given as
\begin{align}
\label{eq:b(0)}
    B_{ia,jb}^{(0)} = 0,
\end{align}
and
\begin{align}
\label{eq:b(1)}
    B_{ia,jb}^{(1)}
    &= \langle i a\vert r_{12}^{-1}\vert j b\rangle,
\end{align}
respectively. The second-order correction has the following form,
\begin{align} 
\label{eq:b(2)}
    B_{ia,jb}^{(2)}
    &= 
    \langle 0 \vert W_N R_0
    [\{i^\dagger a\},[W_N,\{j^\dagger b\}]] 
    \vert 0\rangle \\\notag
    &+
    \langle 0 \vert 
    [\{i^\dagger a\},[W_N,\{j^\dagger b\}]] 
    R_0 W_N
    \vert 0  \rangle \\\notag
    &+
    \langle 0 \vert W_N R_0 
    [\{i^\dagger a\},[H_0,\{j^\dagger b\}]] 
    R_0 W_N\vert 0 \rangle
\end{align}
% \begin{align}
%     B_{ia,jb}^{(2)}
%     &= 
%     \langle \Psi^{(1)} \vert 
%     [\{i^\dagger a\},[W,\{j^\dagger b\}]] 
%     \vert 0\rangle \\\notag
%     &+
%     \langle 0 \vert 
%     [\{i^\dagger a\},[W_N,\{j^\dagger b\}]] 
%     \vert \Psi^{(1)}  \rangle \\\notag
%     &+
%     \langle \Psi^{(1)} \vert 
%     [\{i^\dagger a\},[H_0,\{j^\dagger b\}]] 
%     \vert \Psi^{(1)}  \rangle
% \end{align}
\label{sec:2nd_order}
% \subsection{Equation-of-motion derivation of geminal screened electron-hole interaction kernel}
% \label{sec:fd_gsik}
% \input{sections/004C_gem}
%\subsection{Evaluation of the nested commutators}
\subsection{Rank reduction through commutator evaluation}
The evaluation of the commutator of the excitation operator with the Hamiltonian is needed for simplifying the matrix elements derived in \autoref{sec:pp_def},
\begin{align}
\label{eq:hn_commutator_def}
    [H_N,T_k^\dagger] = \sum_p \epsilon_p [p^\dagger p,i^\dagger a] + \frac{1}{4} \sum_{pqrs}\langle pq \vert r_{12}^{-1} \vert rs \rangle_A[p^\dagger q^\dagger sr,i^\dagger a].
\end{align}
For the one-body operator the commutator can be expressed as,
\begin{align}
\label{eq:one_body_proof}
    [p^\dagger q, \underbrace{x^\dagger y}_B] &= p^\dagger[q,B] + [p^\dagger, B]q \\
    &= p^\dagger \underbrace{[q,x^\dagger y]}_{\delta_{qx}y} + \underbrace{[p^\dagger,x^\dagger y]}_{-\delta_{py}x^\dagger}q \\
    &= \delta_{qx}p^\dagger y - \delta_{py}x^\dagger q,
\end{align}
where the expansion of the commutation product was done using the following commutation identity,
\begin{align}
    [A_1A_2,B] = A_1[A_2,B] + [A_1,B]A_2.
\end{align}
Each commutator is then expressed in terms of anti-commutators using
\begin{align}
    [A,B_1B_2] = [A,B_1]_+B_2 - B_1 [A,B_2]_+ ,
\end{align}
where the anti-commutator is defined as
\begin{align}
    [A,B]_+ = AB + BA.
\end{align}
The anti-commutators in \autoref{eq:one_body_proof} are then evaluated using the fermionic anti-commutator relationship,
\begin{align}
\label{eq:sq_anti_relations}
    [p^\dagger,q^\dagger]_+ = 0 \\
    [p,q]_+ = 0 \\
    [p^\dagger,q]_+ = \delta_{pq}.
\end{align}
Analogously, the commutation of the two-body operator is derived as,
\begin{align}
    [p^\dagger q^\dagger s r,x^\dagger y] = [A_1A_2,B]
\end{align}
where $A_1 = p^\dagger q^\dagger, A_2 = s r,$ and $B = x^\dagger y$.
\begin{align}
    [A_1A_2,B] &= A_1[A_2,B] + [A_1,B]A_2 \\
    &= A_1\underbrace{[s r,x^\dagger y]}_{\delta_{rx}s y - \delta_{sy}x^\dagger r} + \underbrace{[p^\dagger q^\dagger,x^\dagger y]}_{\delta_{qx}p^\dagger y - \delta_{py}x^\dagger q^\dagger}A_2 ,
\end{align}
substituting $A_1$ and $A_2$ back, we get the final expression for the two-body operator,
\begin{align}
\label{eq:two_body_proof}
    [p^\dagger q^\dagger s r,x^\dagger y] = \delta_{rx}p^\dagger q^\dagger s y - \delta_{sy} p^\dagger q^\dagger x^\dagger r + \delta_{qx}p^\dagger y s r - \delta_{py} x^\dagger q^\dagger s r .
\end{align}
An important aspect of both \autoref{eq:one_body_proof} and \autoref{eq:two_body_proof} is that the commutator of one-body and two-body operator with the particle-hole excitation operator is also an one-body and two-body operator, respectively.  
% These results are proven for interpretation of the $T_k^\dagger$ operators in \autoref{sec:2nd_order}. Specifically, \autoref{eq:one_body} and \autoref{eq:two_body} are proven in \autoref{eq:one_body_proof} and \autoref{eq:two_body_proof} respectively. The proof of these results can be found in \renewcommand*{\sectionautorefname}{appendix}\autoref{sec:commutator_proof}.\renewcommand*{\sectionautorefname}{section}
% \begin{align}
% \label{eq:hn_commutator_def}
%     [H_N,T_k^\dagger] = \sum_p \epsilon_p [p^\dagger p,i^\dagger a] + \frac{1}{4} \sum_{pqrs}\langle pq \vert r_{12}^{-1} \vert rs \rangle_A[p^\dagger q^\dagger sr,i^\dagger a]
% \end{align}
% which are commutators with respect to the one-body operator $H_0$ and the two-body operator $W_N$ respectively.
\subsection{Diagrammatic evaluation of the matrix elements}
\label{sec:diagrams}
\begin{figure}[h!]
\centering
\includegraphics[scale=0.3]{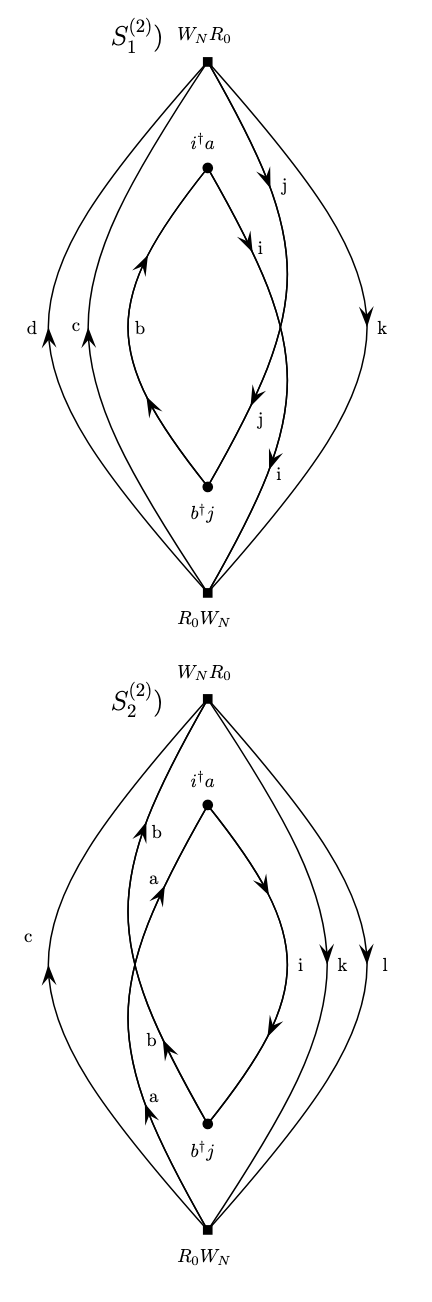}
\includegraphics[scale=0.3]{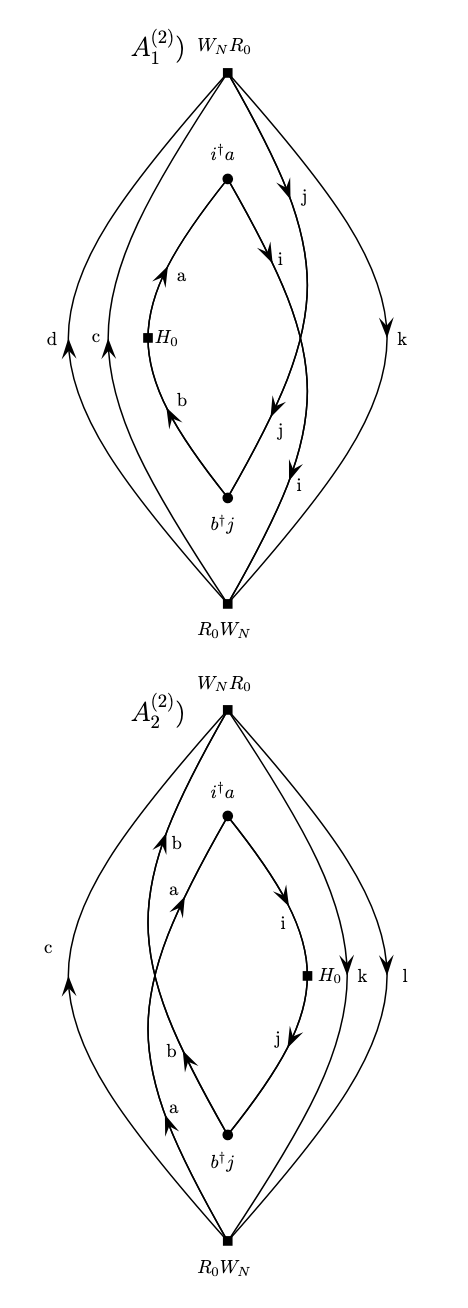}
\includegraphics[scale=0.3]{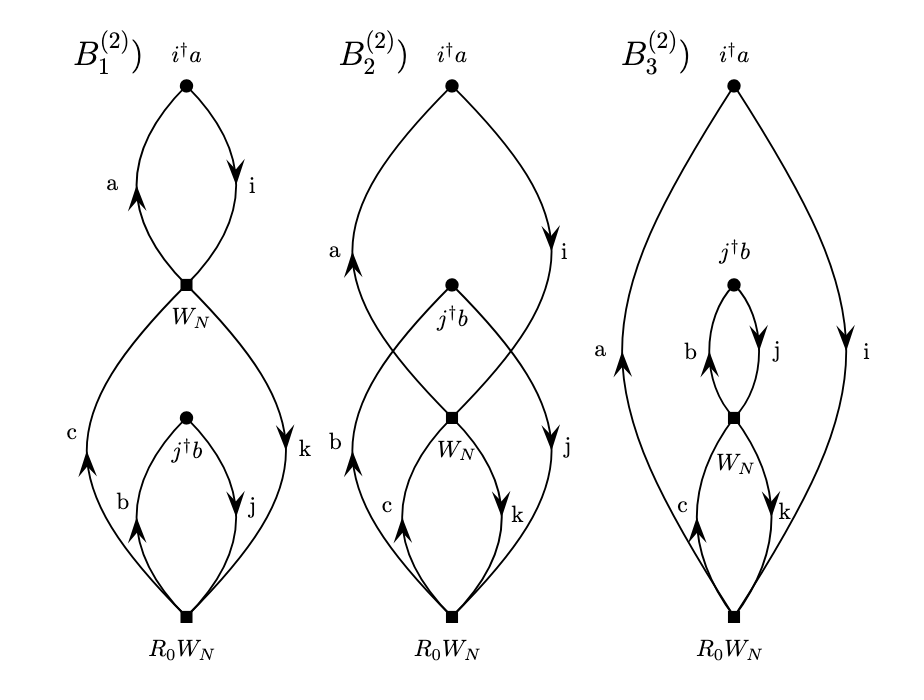}
\caption{\centering The contributing terms of the $\mathbf{M}$ matrix in terms of Hugenholz diagrams. Each diagram is labeled with the piece of the $\mathbf{M}$ matrix that it comes from ($\mathbf{A}$, $\mathbf{B}$, or $\mathbf{S}$). These represent the expectation values required to calculate the $\mathbf{M}$ operator up to $2^{\mathrm{nd}}$ order.}
\label{fig:hugenholz_diags}
\end{figure}

% \begin{align}
%     W_N 
%     = 
%     \frac{1}{4} \sum_{pqrs} \langle pq \vert r_{12}^{-1} \vert rs \rangle_A 
%     \{p^\dagger q^\dagger s r\}
% \end{align}
% \begin{align}
%     [W,T] = [W_N,T]
% \end{align}
% \begin{align}
%     G_N 
%     = 
%     \frac{1}{4} \sum_{pqrs} \langle pq \vert g \vert rs \rangle_A 
%     \{p^\dagger q^\dagger s r\}
% \end{align}
% \begin{align}
%     \langle 0 \vert \{i^\dagger a\}[W_N,\{a^\dagger i\}]\vert 0  \rangle
%     &=
%     \langle 0 \vert \{i^\dagger a\}W_N\{a^\dagger i\}\vert 0  \rangle_\mathrm{linked} \\\notag
%     &-\langle 0 \vert \{i^\dagger a\}\{a^\dagger i\}W_N\vert 0  \rangle_\mathrm{linked} \\\notag
%     \\\notag
%     &=
%     D_2
% \end{align}
% \begin{align}
%     \langle 0 \vert \{i^\dagger a\} 
%     [W_N,\{a^\dagger i\}] G_N \vert 0 \rangle
%     &=
%     \langle 0 \vert \{i^\dagger a\} 
%     W_N \{a^\dagger i\} G_N \vert 0 \rangle_\mathrm{linked} \\\notag
%     &-\langle 0 \vert \{i^\dagger a\} 
%     \{a^\dagger i\} W_N G_N \vert 0 \rangle_\mathrm{linked} \\\notag
%     \\\notag
%     &= D_3 - D_4
% \end{align}
% %%%%%%%%%%%%%%%%%%%%%%%%%%%%%%%%%%%
% %%%%%%%%%%%%%%%%%%%%%%%%%%%%%%%%%%%
We note that all the matrix elements of of the $\bm{S}$, $\bm{A}$, and $\bm{B}$ matrices can be expressed as expectation values with respect to the Fermi vacuum and has the general structure $\langle 0 \vert ... \vert 0 \rangle$. Because of this form they are especially convenient to be solved using diagrammatic techniques and only fully connection diagrams will have non-zero contributions to the overall expression. 
This implies that the non-zero terms in \autoref{eq:s(2)}, \autoref{eq:a(2)}, and \autoref{eq:b(2)} can be evaluated as 
\begin{align}
\label{eq:s_dia}
    S_{ia,jb}^{(2)}
    &=
    \langle 0 \vert W_N R_0  \{i^\dagger a\} \{b^\dagger j\} R_0 W_N \vert 0\rangle_\mathrm{linked}
\end{align}
\begin{align}
\label{eq:a_dia}
    A_{ia,jb}^{(2)}
    &=
    \langle 0 \vert W_N R_0 
    [\{i^\dagger a\},[H_0,\{b^\dagger j\}]] 
    R_0 W_N\vert 0\rangle_\mathrm{linked}    
\end{align}
\begin{align} 
\label{eq:b_dia}
    B_{ia,jb}^{(2)}
    &= 
    \langle 0 \vert W_N R_0
    [\{i^\dagger a\},[W_N,\{j^\dagger b\}]] 
    \vert 0\rangle \\\notag
    &+
    \langle 0 \vert 
    [\{i^\dagger a\},[W_N,\{j^\dagger b\}]] 
    R_0 W_N
    \vert 0  \rangle \\\notag
    &+
    \langle 0 \vert W_N R_0 
    [\{i^\dagger a\},[H_0,\{j^\dagger b\}]] 
    R_0 W_N\vert 0 \rangle_\mathrm{linked}
\end{align}
Additionally, further simplification can be achieved by noting that the operation of $R_0 W_N$ on the Fermi-vacuum $\vert 0 \rangle$ will only generate 2p-2h transitions. 
\begin{align}
    R_0 W_N \vert 0 \rangle 
    = \frac{1}{4} \sum_{ijab} [R_0 W_N]_{abij}\{b^\dagger a^\dagger i j\}
\end{align}
The non-zero contributing Hugenholtz diagrms for the three terms are presented in \autoref{fig:hugenholz_diags}.
\subsection{Evaluation of interior eigenvalues using inverted Krylov basis and folded-spectrum}
\label{sec:krylov_theory}
One of the central challenges in the calculation of the excitation energies of quantum dots is the large size of matrices. As shown in \autoref{tab:amat_size}, the 1p-1h matrix size is large even for small clusters.
One strategy to reduce the computational cost is to realize that, for the polarization propagator calculation, not all excitation energies are equally important or needed. Specifically, for a given values of $\omega$, the sum is dominated by the excitation in the vicinity of the poles $(\omega_{0k}-\omega)^{-1}$. Therefore, in this work, we do not aim to evaluate the entire eigenspectrum of the $\mathbf{M}$ matrix; instead we only calculate a subset of eigenvalues and eigenvectors in the vicinity of a given value of $\omega$. Calculation of the interior eigenvalue problem has been a recurring theme for excited state calculations, and various techniques such as filter-diagonalization\cite{Yu1999filterdiag,Wall1995filterdiag,Pieper2016filterdiag}, polynomial-based energy filtering, and folded-spectrum based methods\cite{Wang1994foldedspec,Canning2000foldedspec} have been used to calculate selected eigenvalues. We choose to use the folded spectrum method for this work, which begins with the generalized eigenvalue problem
\begin{align}
    \mathbf{A}\mathbf{x} = \omega \mathbf{B}\mathbf{x}.
\end{align}
This is equivalent to the form of the M-operator in \autoref{eq:m_eigen}. Using the notation from \autoref{eq:m_eigen} we rewrite the generalized eigenvalue problem as
\begin{align}
    \mathbf{M}\mathbf{x} = \omega \mathbf{S}\mathbf{x},
\end{align}
which can be rewritten as
\begin{align}
    (\mathbf{M} - \omega \mathbf{S})\mathbf{x} = 0. 
\end{align}
Thus, this problem becomes calculation of the null space of $(\mathbf{M} - \omega \mathbf{S})$, which can be done using the folded spectrum method to "fold" the eigenspectrum of $(\mathbf{M} - \omega \mathbf{S})$ around a given $\omega$ with 
\begin{align}
    (\mathbf{M} - \omega \mathbf{S})^2 \mathbf{x} = 0
\end{align}
then, finding the null space of this folded operator is identical to unconstrained minimization of the Rayleigh coefficients
\begin{align}
    \rho_\mathrm{\omega}(\mathbf{y}) = \langle \mathbf{y} \vert (\mathbf{M} - \omega \mathbf{S})^2 \vert \mathbf{y} \rangle
\end{align}
\begin{align}
    \mathbf{x} =  \underset{\mathbf{y}}{\mathrm{argmin}}\, \rho_\mathrm{\omega}(\mathbf{y}).
\end{align}
\par
In general, this method is computationally intractable for large basis sets such as those listed in \autoref{tab:amat_size}, since the squared operator is computationally expensive. Additionally, for the PP we are not seeking the eigenvalues of $(\mathbf{M} - \omega \mathbf{S})$, but rather the resolvent operator $(\mathbf{M} - \omega \mathbf{S})^{-1}$ which adds a large amount of computational overhead via matrix inversion. To use the folded spectrum method then, we focus our attention on both reducing the size of the basis while still maintaining accuracy in our eigenvalue calculations, and avoiding explicit inversion of the resolvent operator.
%%%%%%%%%%%%%%%%%%%%%%%%%%%%%%%%%%%%%%%%%%%%%%%%%%%%%%%%%%%%%%%%%%%%%%%%%
\subsubsection{Subspace projection using inverted Krylov basis}
A Krylov subspace of $m^{\mathrm{th}}$ order is defined as
\begin{align}
\label{eq:krylov_def}
    \mathcal{K}_m = \mathrm{span}(v,\mathbf{A}v,\mathbf{A}^2v,\ldots,\mathbf{A}^{m-1}v)
\end{align}
where $\mathbf{A} \in \mathbb{R}^{n \times n}$ is the $n \times n$ matrix to be projected into the $m^{\mathrm{th}}$ order subspace, and $v \in \mathbb{R}^n$ a random n-dimensional vector. This subspace corresponds to repeated applications of the $\mathbf{A}$ operator, which enriches the largest eigenvalues of $\mathbf{A}$ via power iteration. For the polarization propagator, however, we require access to the eigenvalues of the resolvent operator
\begin{align}
    \mathbf{R} = (\mathbf{M} - \omega \mathbf{S})^{-1}
\end{align}
which becomes an inverted Krylov subspace when applied using the definition in \autoref{eq:krylov_def}. This inverse subspace is defined as
\begin{align}
    \mathcal{K}_m = \mathrm{span}(v,\mathbf{R}v,\mathbf{R}^2v,\ldots,\mathbf{R}^{m-1}v)
\end{align}
which, just as in \autoref{eq:krylov_def} corresponds to repeated applications of the resolvent operator to a guess vector $v$. This is equivalent to an inverse power iteration centered on the shift $\omega$. Expanding a vector $x$ in the eigenbasis $\{v_i\}$ of $\mathbf{M}$ with eigenvalues $\{\lambda_i\}$
\begin{align}
\label{eq:krylov_basis_expansion}
    (\mathbf{M}-\omega\mathbf{S})^{-m}\mathbf{x} = \sum_{i=1}^N \frac{1}{(\lambda_i - \omega)^m} c_i v_i,
\end{align}
demonstrates that components with $\lambda_i$ close to $\omega$ are preferentially amplified as $m$ increases. Thus the inverse Krylov subspace selectively enriches the subspace associated with the eigenvalues near the chosen shift $\mu$. Once again, the operator of interest in this work is not a standard operator but a superoperator that encodes excitation energies of the systems, further complicating the generated Krylov subspace.
This approach is connected to performing a finite-order real-space approximation of the Riesz projector discussed in \autoref{sec:riesz_proj}. While the Riesz projector is able to contour around the exact eigenvalues of the resolvent, the inverse Krylov subspace iteratively enriches the eigenvalues, meaning there is some numerical error at low values of $m$ in $\mathcal{K}_m$. By forming our subspace using this iterative approach we not only project the full Hamiltonian into a more computationally efficient space, but our basis rotates as $\omega$ changes. Note that due to the resolvent operator having $\omega$ dependence, the basis formed also depends on $\omega$, and therefore by constructing this basis we are guaranteed for the guess vectors $x$ and $y$ to be enriched around the pole where $\omega_{k0} = \omega$. This is equivalent to letting the basis vectors $x$ be formed as
\begin{align}
    \mathbf{x} = \sum_{i=1}^n \mathbf{y}(\omega)
\end{align}
and using the linear combination of frequency-dependent $y$ vectors to apply the folded spectrum method on a set of basis functions iteratively enriched to include information about the poles of $\mathbf{R}(\omega)$. 
%%%%%%%%%%%%%%%%%%%%%%%%%%%%%%%%%%%%%%%%%%%%%%%%%%%%%%%%%%%%%%%%%%%%%%%%%%
\subsubsection{Approximate inversion using weighted Jacobi iteration}
To obtain the inverse Krylov basis we have to solve the following linear equation
\begin{align}
    (\mathbf{M} - \omega \mathbf{S}) \mathbf{v}^{(k+1)} = \mathbf{v}^{(k)}
\end{align}
and we use the Jacobi iteration to obtain the approximate inverse of $(\mathbf{M} - \omega \mathbf{S})$.
Splitting the matrix into diagonal $(\mathbf{D})$ and off-diagonal components $(\mathbf{F})$,
\begin{align}
    (\mathbf{M}-\omega \mathbf{S}) = \mathbf{D} + \mathbf{F}, 
\end{align}
The Jacobi vector is defined as,
\begin{align}
    \mathbf{v}^{(k+1)}_\mathrm{Jaco} = \mathbf{D}^{-1} \mathbf{v}^{(k)} - \mathbf{D}^{-1}\mathbf{F}\mathbf{v}^{(k)}.
\end{align}
However, this process is not guaranteed to converge if the spectral radius of $- \mathbf{D}^{-1}\mathbf{F}$
is greater than one. To address this issue, we use a weighted Jacobi approach where the new vector is constructed from the average of the old vector and the Jacobi vector
\begin{align}
    \mathbf{v}^{(k+1)} = \mathbf{v}^{(k)} + \alpha[\mathbf{v}^{(k+1)}_\mathrm{Jaco}-\mathbf{v}^{(k)}],
\end{align}
where $\alpha$ is the weighting parameter.
Because this is part of an inverse Krylov subspace construction process and we expect subsequent basis vectors to be enriched in the lowest eigenvalue, we choose $\alpha$ by minimizing the following Rayleigh coefficient
\begin{align}
    \rho^{(k+1)} = \frac{[\mathbf{v}^{(k+1)}]^T \mathbf{Z} \mathbf{v}^{(k+1)}}{[\mathbf{v}^{(k+1)}]^T \mathbf{v}^{(k+1)}}.
\end{align}
The above has the general form,
\begin{align}
    \rho^{(k+1)} = \frac{\rho^{(k)} + a_1 \alpha + a_2 \alpha^2}{1 + b_1 \alpha + b_2 \alpha^2}
\end{align}
and was used to optimize $\alpha$ such that $\rho^{(k+1)} < \rho^{(k)}$.

\section{Results}
\label{sec:results}

\begin{table}[h]
\centering
\begin{ruledtabular}
\caption{Location and intensity of peaks for the PbS quantum dots, at each level of correlation.}
\label{tab:max_peaks_pbs}
\begin{tabular}{c c c c} 

 \rule[-1ex]{0pt}{2.5ex} Chemical System  & Order of correction & Peak Location (eV) & Peak Intensity\\
 \hline
 \ce{Pb4S4} & $0^{\mathrm{th}}$ order & 9.85 & 13.49\\
 \ce{Pb4S4} & $1^{\mathrm{st}}$ order & 9.30 & 9.38\\
 \ce{Pb4S4} & $2^{\mathrm{nd}}$ order & 9.30 & 9.21\\
 
 \ce{Pb44S44} & $0^{\mathrm{th}}$ order & 9.35 & 1.09\\
 \ce{Pb44S44} & $1^{\mathrm{st}}$ order & 9.05 & 5.58\\
 \ce{Pb44S44} & $2^{\mathrm{nd}}$ order & 5.89 & 7.24\\
 
 \ce{Pb140S140} & $0^{\mathrm{th}}$ order & 9.55 & 0.71\\
 \ce{Pb140S140} & $1^{\mathrm{st}}$ order & 9.65 & 11.82\\
 \ce{Pb140S140} & $2^{\mathrm{nd}}$ order & 9.65 & 11.17\\

\end{tabular}
\end{ruledtabular}
\end{table}

\begin{table}[h]
\centering
\begin{ruledtabular}
\caption{Location and intensity of peaks for the CdS quantum dots, at each level of correlation.}
\label{tab:max_peaks_cds}
\begin{tabular}{c c c c} 

 \rule[-1ex]{0pt}{2.5ex} Chemical System  & Order of correction & Peak Location (eV) & Peak Intensity\\
 \hline
 
 \ce{Cd24S24} & $0^{\mathrm{th}}$ order & 5.35 & 2.66\\
 \ce{Cd24S24} & $1^{\mathrm{st}}$ order & 6.90 & 9.14\\
 \ce{Cd24S24} & $2^{\mathrm{nd}}$ order & 6.90 & 9.84\\
 
 \ce{Cd45S45} & $0^{\mathrm{th}}$ order &  9.35 & 1.02\\
 \ce{Cd45S45} & $1^{\mathrm{st}}$ order & 8.10 & 10.66\\
 \ce{Cd45S45} & $2^{\mathrm{nd}}$ order & 8.10 & 10.94\\
 
\end{tabular}
\end{ruledtabular}
\end{table}

Using the derivation presented in \autoref{sec:theory}, the electronic spectra were computed for two classes of semiconductor quantum dots: lead selenide (PbS) and cadmium selenide (CdS). Calculations were carried out on the following cluster: Pb\textsubscript{4}S\textsubscript{4}, Pb\textsubscript{44}S\textsubscript{44}, Pb\textsubscript{140}S\textsubscript{140}, Cd\textsubscript{24}S\textsubscript{24}, and Cd\textsubscript{45}S\textsubscript{45}.
The structures of these quantum dots were obtained from their respective crystal structures, and single particle states from each system were obtained from Hartree-Fock calculations using the software TERACHEM\cite{TeraChem2020}. The single particle states were used for the reference state and reference energies, before corrections were applied. To apply these corrections, we used the density-fitting approach for efficient evaluation of the two-electron integrals.
% The method described above was used to calculate the electronic specta of two sets of semiconductor quantum dots, Lead Selenide (PbS) and Cadmium Selenide (CdS).
% We evaluate the polarization propagator using the resolvent formulation described in \autoref{eq:full_pp_resolvent}, applying the results to a set of lead selenide (PbS) and cadmium selenide (CdS) semiconductor quantum dots: Pb\textsubscript{4}S\textsubscript{4}, Pb\textsubscript{44}S\textsubscript{44}, Pb\textsubscript{140}S\textsubscript{140}, Cd\textsubscript{24}S\textsubscript{24}, and Cd\textsubscript{45}S\textsubscript{45}. 
All spectra were computed on a common frequency range and plotted as absorption-like response functions. The Hartree-Fock (HF) spectra is shown for each system as a zeroth order comparison. The HF spectra was calculated by taking the zeroth order approximation of the M-operator described in \autoref{eq:m_eigen}, which involves setting $\mathbf{A}=A^{(0)}_{ia,jb} = (\epsilon_a - \epsilon_i)\delta_{ij}\delta{ab}$, $\mathbf{S}=S^{(0)}_{ia,jb} = \delta_{ij}\delta_{ab}$, and $\mathbf{B}=B^{(0)}_{ia,jb} = 0$. In this approximation the M-operator is diagonal and easily invertible for use in the polarization propagator formula. In \autoref{sec:effect_of_size} we discuss show how changing the diameter of the quantum dots affects the HF spectra, and the $1^{\mathrm{st}}$ order corrected spectra, while in \autoref{sec:effect_of_correlation} we specifically investigate the effect that including $1^{\mathrm{st}}$ order correlated effects has on the spectra, and higher order effects are shown in \autoref{sec:ho_effects}. All results for each order of correction are presented in \autoref{tab:max_peaks_pbs} and \autoref{tab:max_peaks_cds}.
\subsection{Effect of size on optical absorption spectra}
\label{sec:effect_of_size}
\begin{figure}[h]
\centering
\includegraphics[width=0.45\textwidth]{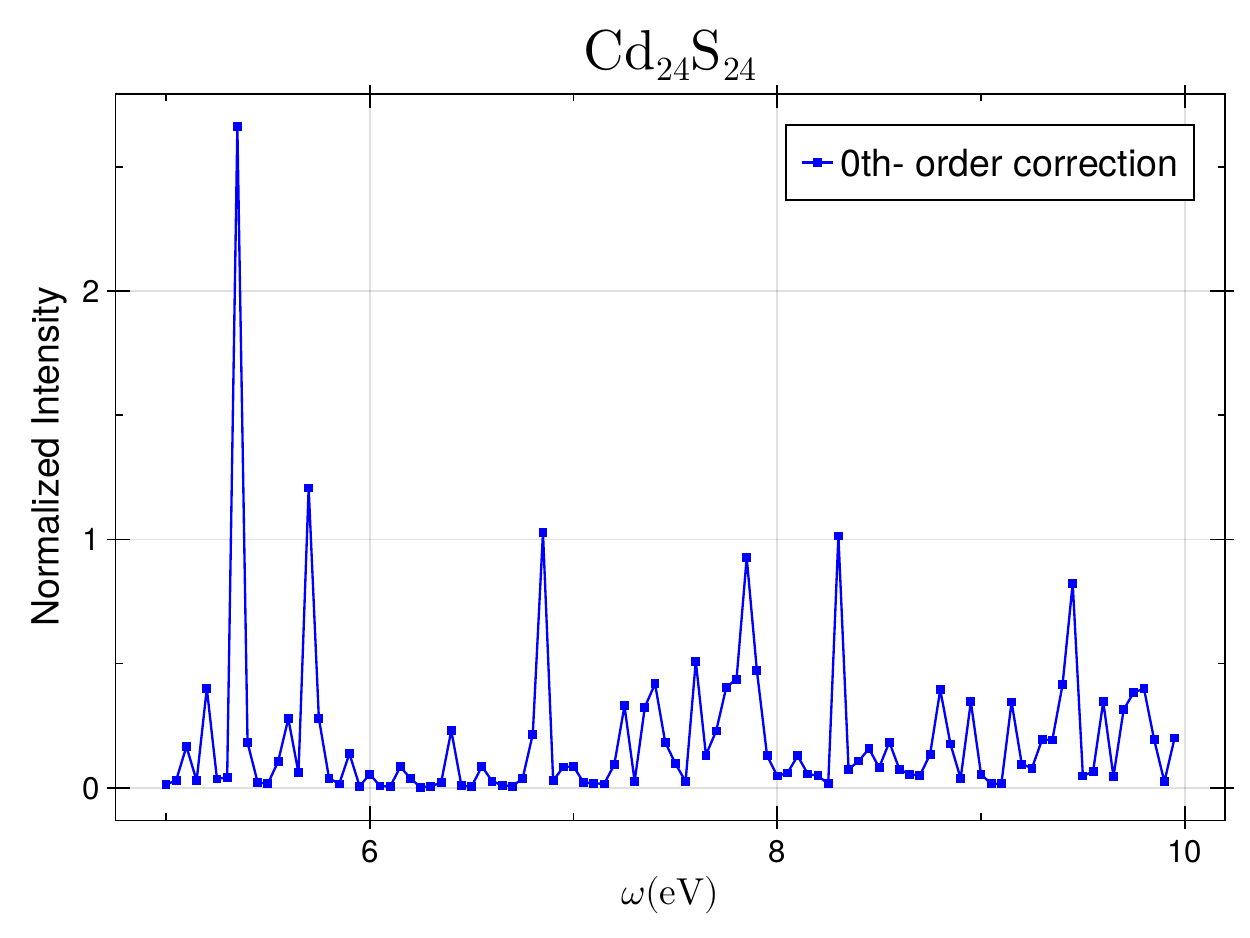}
\includegraphics[width=0.45\textwidth]{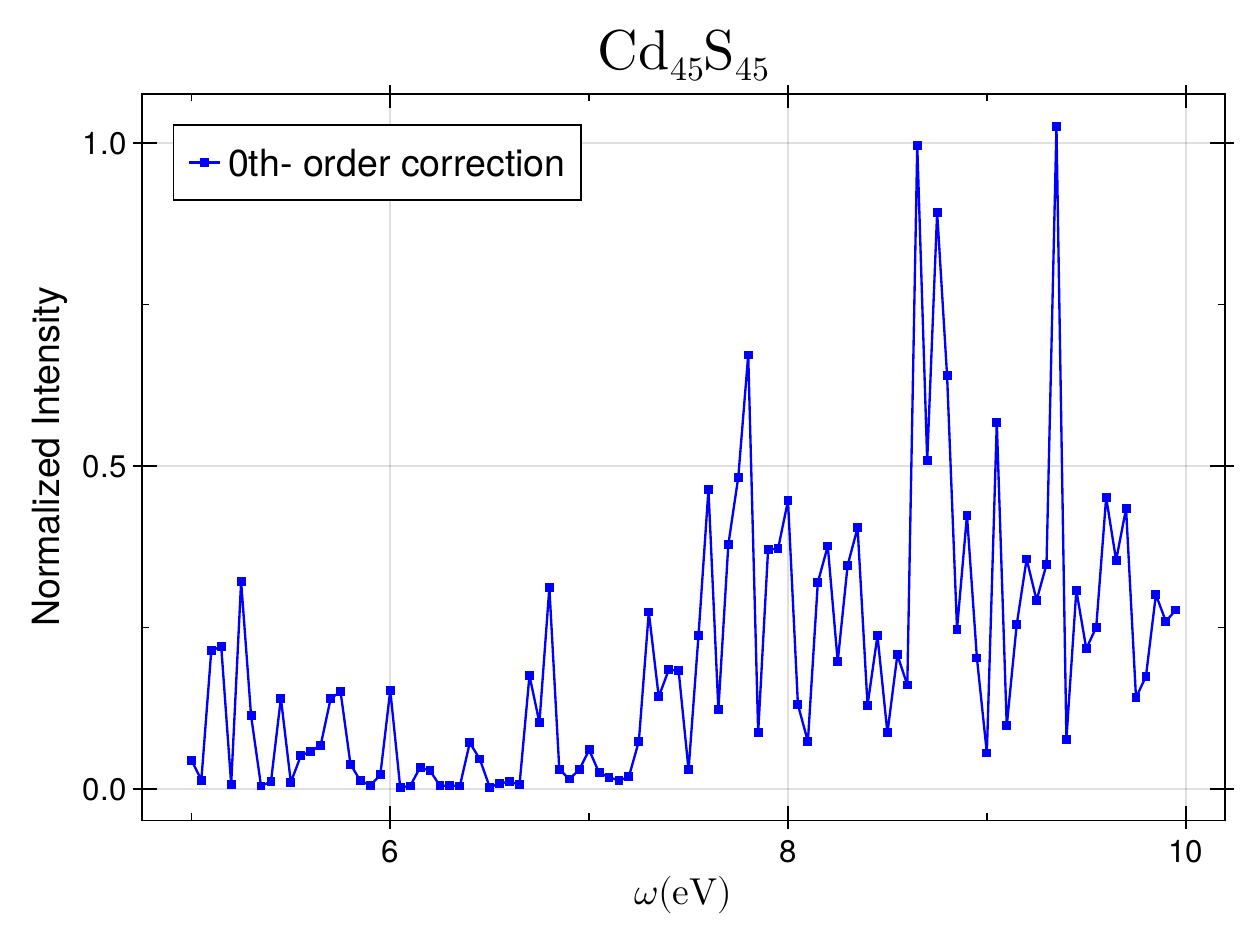}
\caption{\centering Absorption spectra of CdS QDs from the zeroth order HF approximation of the M-operator}
\label{fig:cds_hf}
\end{figure}

\begin{figure}[h]
\centering
\includegraphics[width=0.3\textwidth]{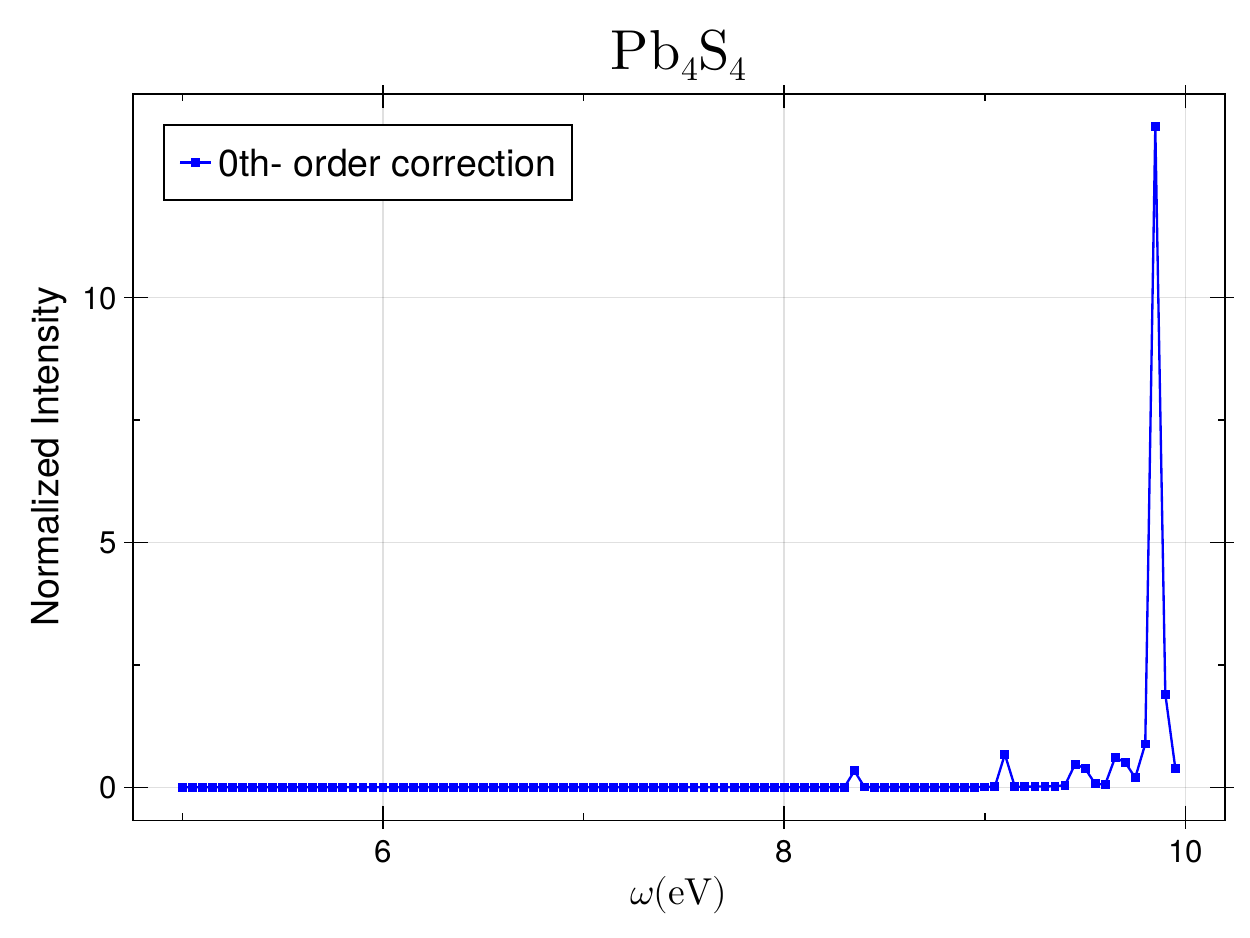}
\includegraphics[width=0.3\textwidth]{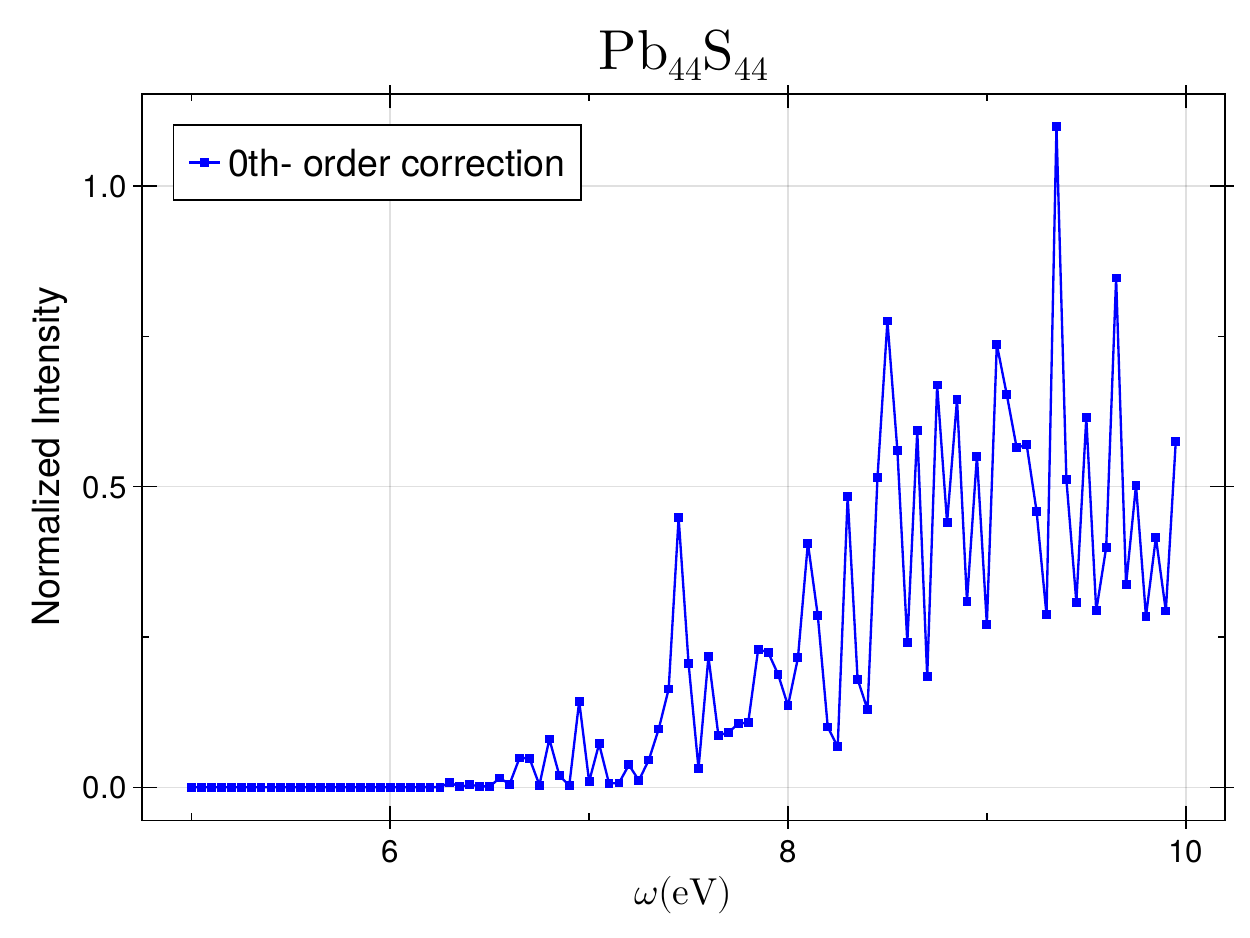}
\includegraphics[width=0.3\textwidth]{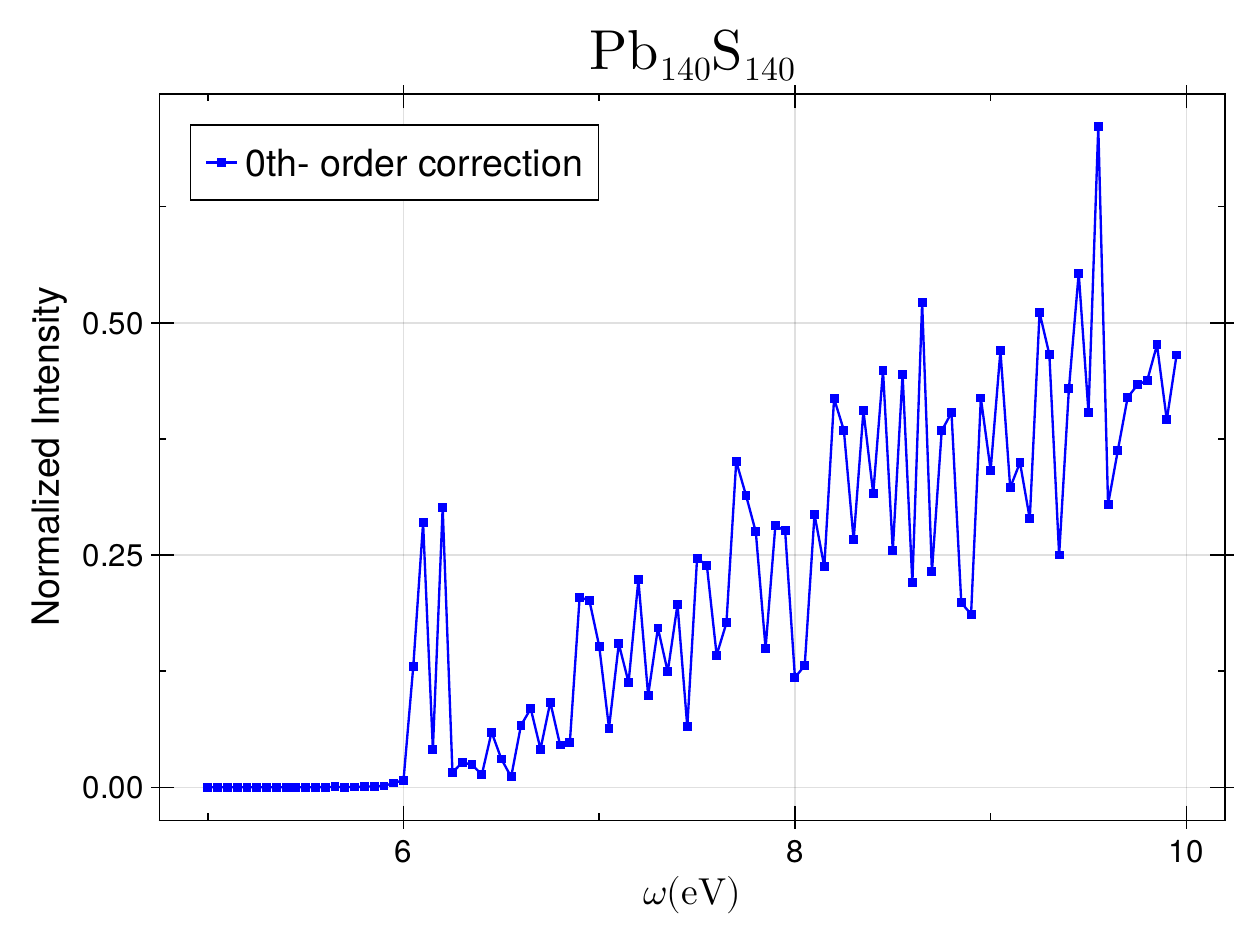}
\caption{\centering Absorption spectra of PbS QDs from the zeroth order HF approximation of the M-operator}
\label{fig:pbs_hf}
\end{figure}

One of the primary properties that inspired investigation into quantum dots was the strong dependence on system size to light-matter interaction\cite{Brus1984size,Efros1998size,Moreels2009size}. This effect is seen strongly in the systems described in \autoref{tab:length_scale}. The zeroth order HF spectra can be seen in \autoref{fig:pbs_hf} and \autoref{fig:cds_hf}.
In the CdS system, we can see that the increasing system size ($\ce{Cd_{24}S_{24}} \rightarrow \ce{Cd_{45}S_{45}}$) results in a blue shifting of the zeroth order absorption lines. The maximum intensity transition shifts 4 eV from 5.35 eV to 9.35 eV (see \autoref{tab:max_peaks_cds}). However, the size increase also results in a lower overall maximum probability transition, a 62\% decrease in intensity, which is also shown in \autoref{tab:max_peaks_cds}.
\begin{table}[!h]
\centering
\begin{ruledtabular}
\caption{Length scale as a function of diameter for each dot.}
\label{tab:length_scale}
\begin{tabular}{c c c} 

 \rule[-1ex]{0pt}{2.5ex} Chemical System  & Diameter (AU) & Avg Separation (AU) \\
 \hline
 \ce{Pb4S4} & 2.97 & 1.65\\
  
 \ce{Pb44S44} & 14.84 & 8.24\\
 
 \ce{Pb140S140} & 39.26 & 21.78 \\
 
 \ce{Cd24S24} & 20.09 & 11.16\\
 
 \ce{Cd45S45} & 24.32 & 13.48\\
 
\end{tabular}
\end{ruledtabular}
\end{table}

The PbS system has a different response to increasing system size. The smallest system, $\ce{Pb_4S_4}$ has a single dominant peak at 9.85 eV, and the peak remains in both the $\ce{Pb_{44}S_{44}}$ and $\ce{Pb_{140}S_{140}}$. However, due to the rapidly increasing size of the state space more transitions are made available in the larger systems, including the introduction of strong low energy transitions in the $\ce{Pb_{140}S_{140}}$ system. These results show how uncorrelated systems behave as a function of system size, but with the M-operator formulation we can also investigate how impact of inclusion of electron-electron correlation impact this behavior.
\begin{figure}[h]
\centering
\includegraphics[width=0.3\textwidth]{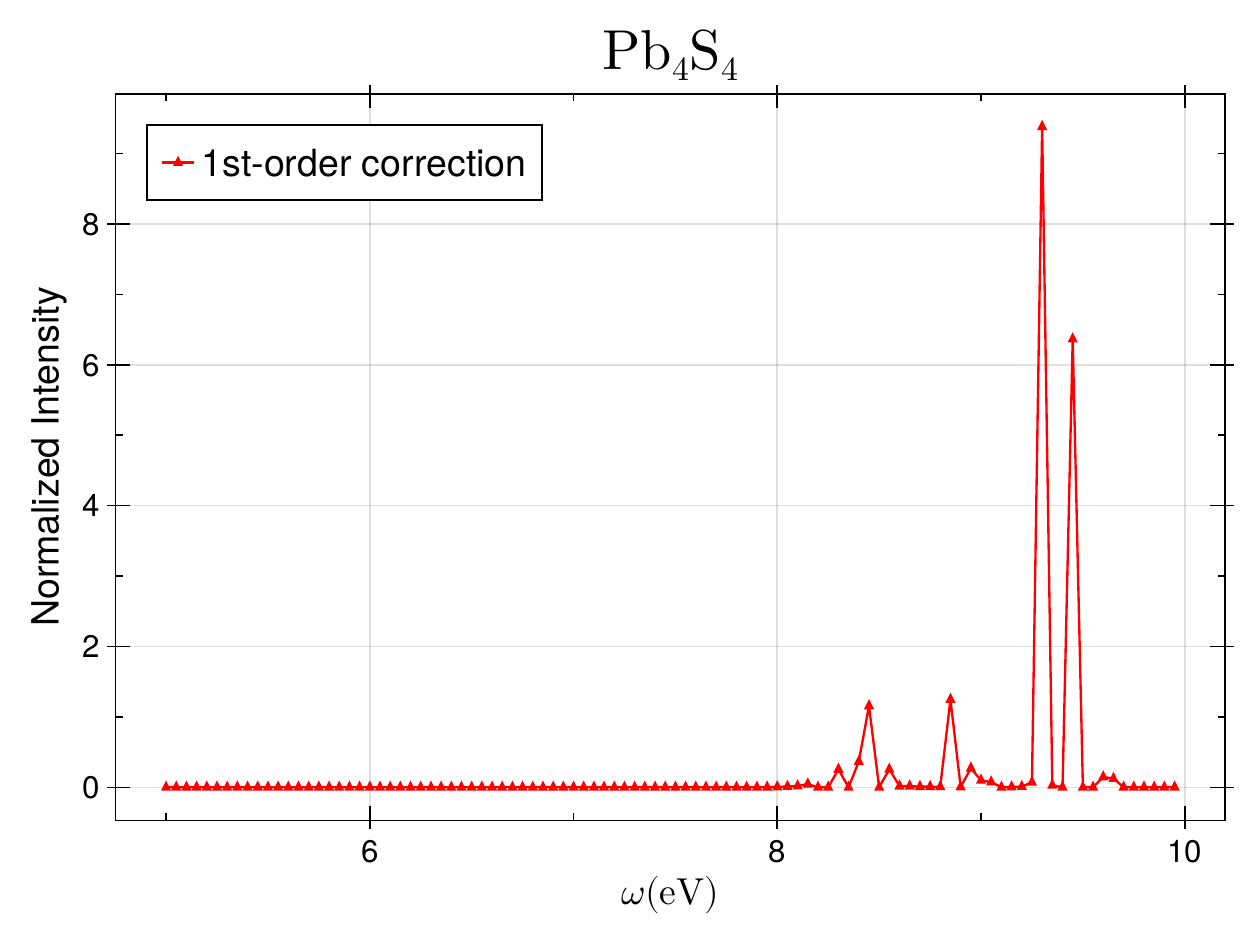}
\includegraphics[width=0.3\textwidth]{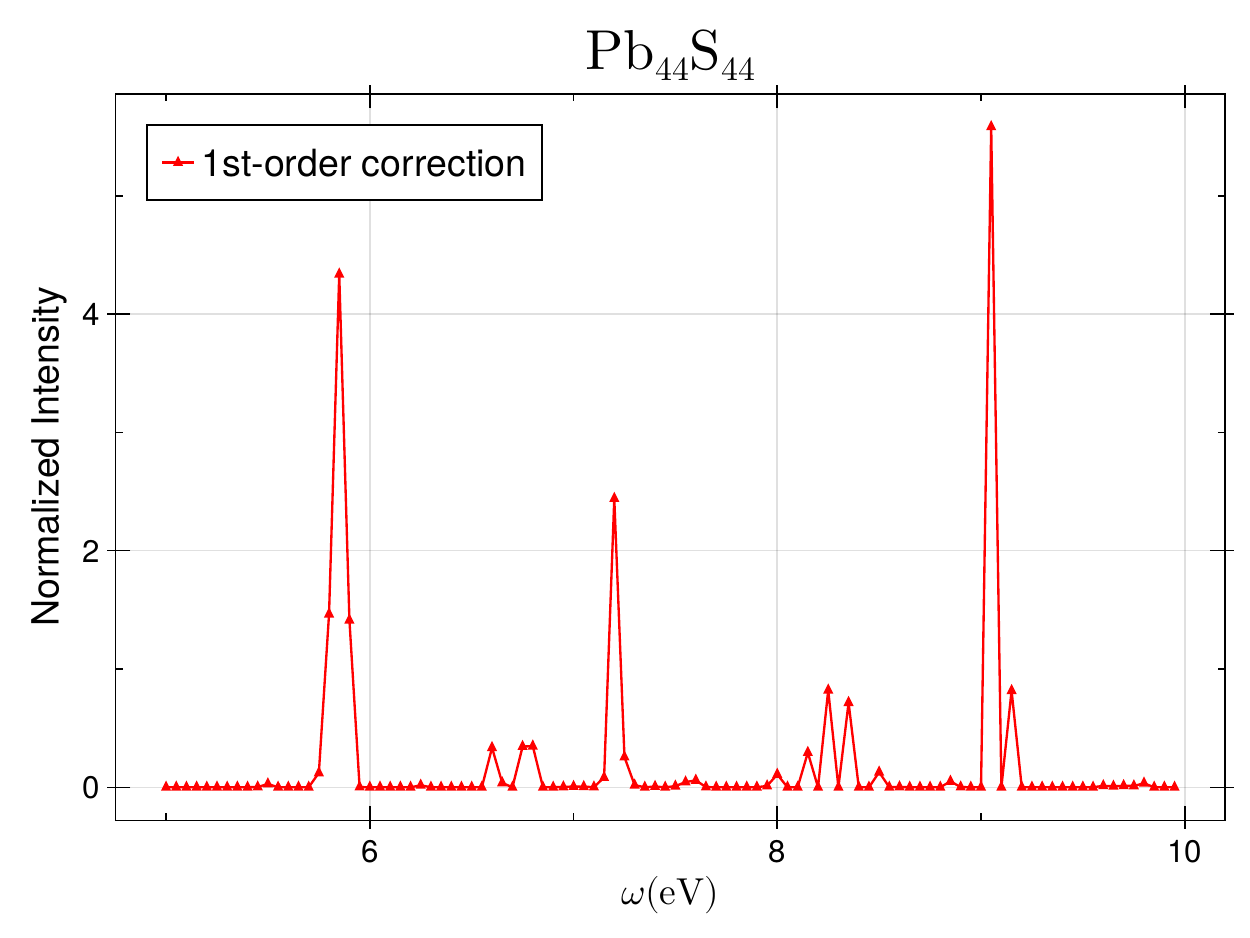}
\includegraphics[width=0.3\textwidth]{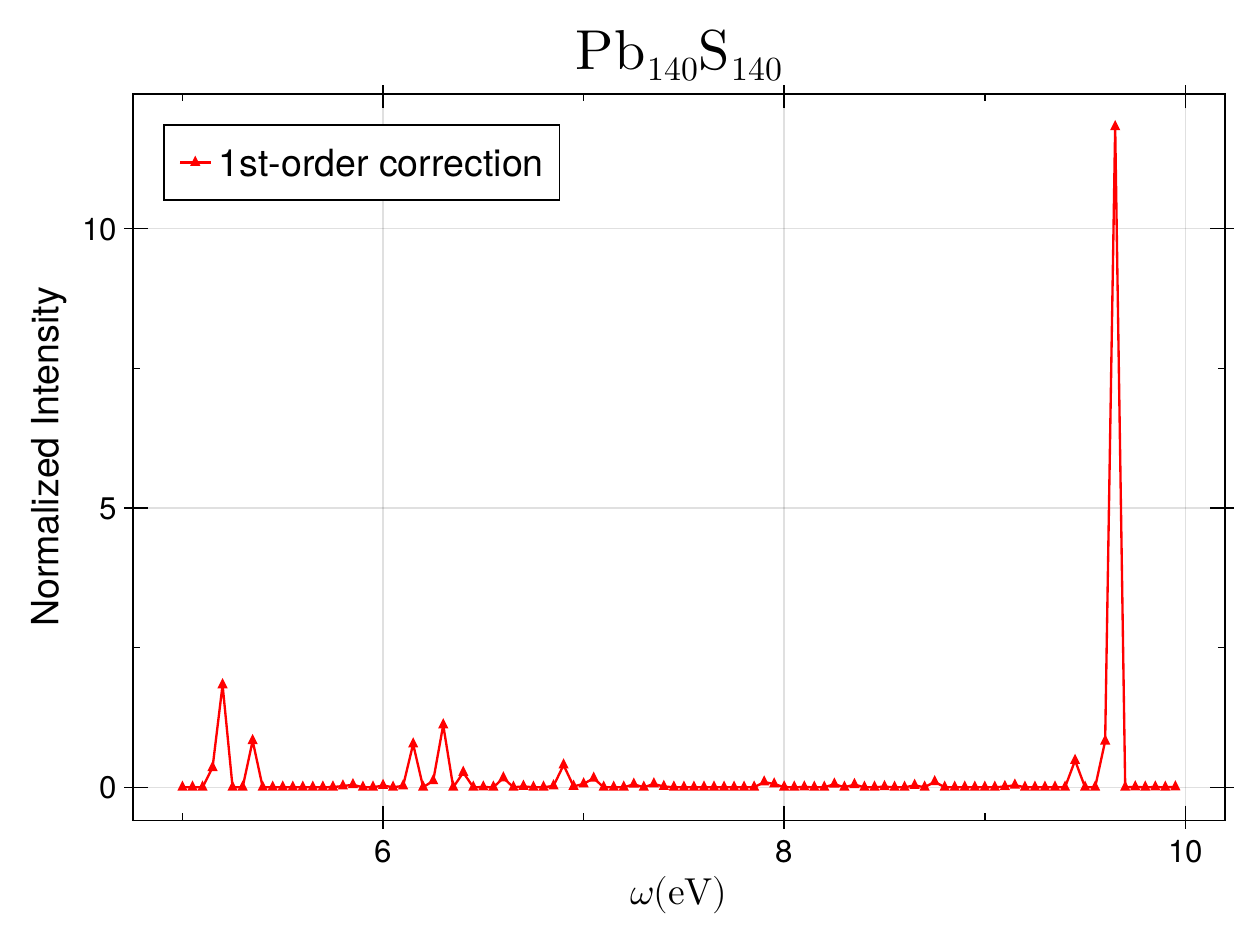}
\caption{\centering Absorption spectra of PbS QDs from the $1^{\mathrm{st}}$ order correction to the approximation of the M-operator}
\label{fig:pbs_1st}
\end{figure} 
\newpage
\begin{figure}[h]
\centering
\includegraphics[width=0.45\textwidth]{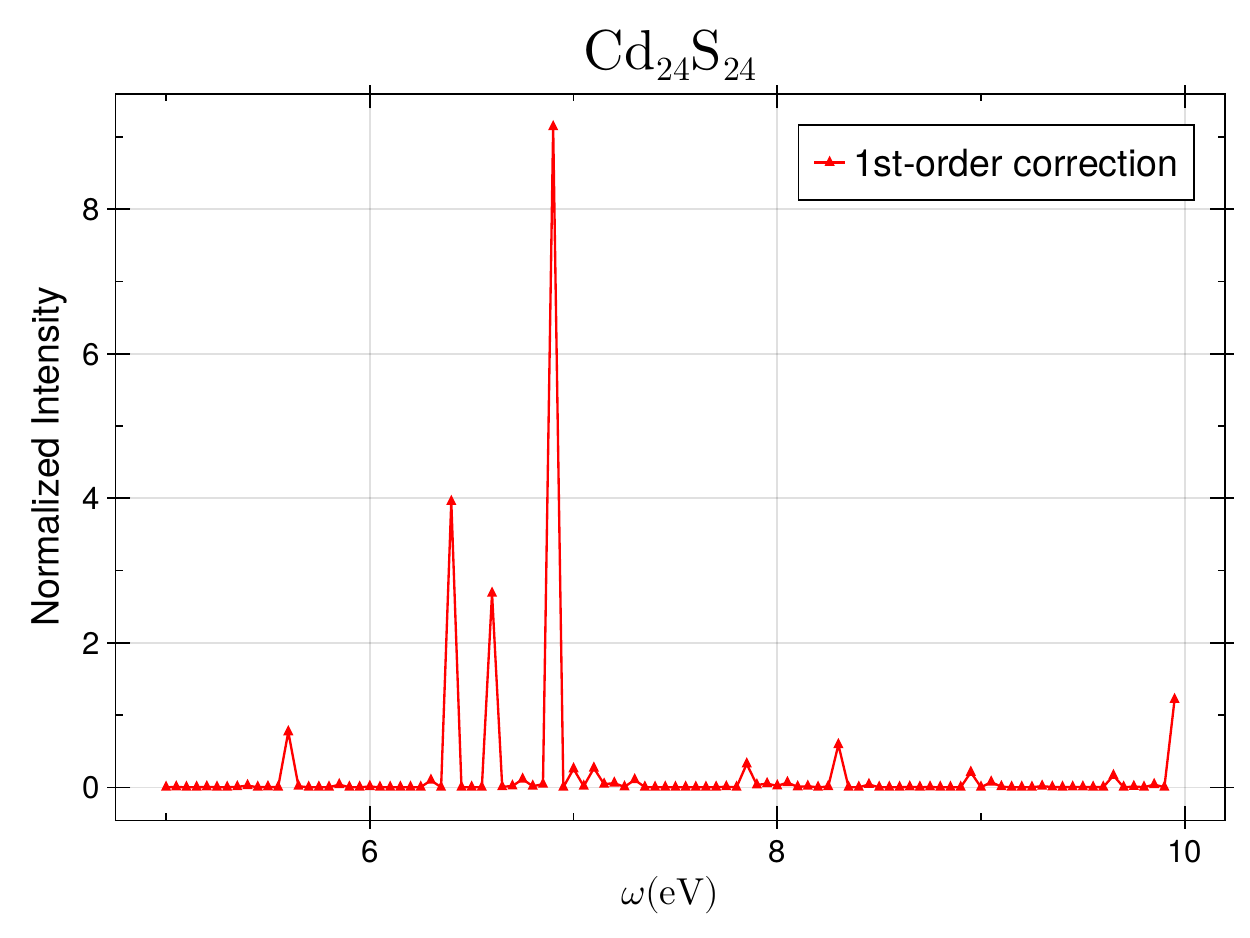}
\includegraphics[width=0.45\textwidth]{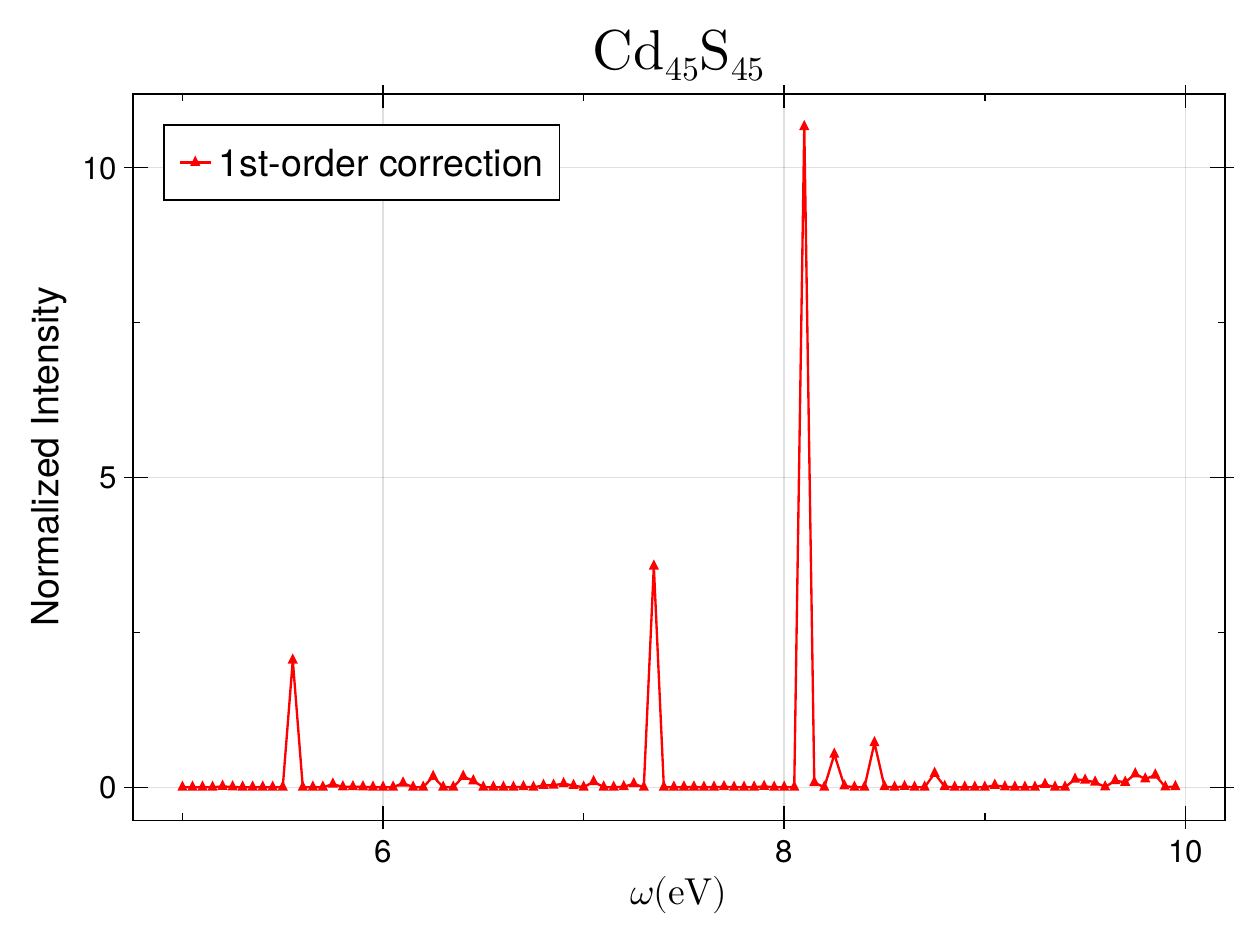}
\caption{\centering Absorption spectra of CdS QDs from the $1^{\mathrm{st}}$ order correction to the approximation of the M-operator}
\label{fig:cds_1st}
\end{figure} 
The correlated spectra have new properties compared to their uncorrelated counterparts. When looking at the CdS system we again see the blue shifting of the dominant transition (6.90 eV $\rightarrow$ 8.10 eV) but we also see transitions in the 6-7 eV range disappear, with new transitions <6 eV and >7eV showing up. This indicates that the size of the correlated system has a strong effect on the absorption spectra in CdS quantum dots. A similar trend is observed in the PbS system where the high energy transitions in $\ce{Pb_4S_4}$ are replaced in $\ce{Pb_{44}S_{44}}$ with a much broader range of possible transitions. Interestingly, continuing the increase of system size leads to an absorption spectra similar to $\ce{Pb_4S_4}$, where the $\ce{Pb_{140}S_{140}}$ system has a dominant high energy transition, though lower energy transitions are possible in this configuration. These results show that the interaction properties of both Pbs and CdS quantum dots have a strong dependence on system size.
\subsection{Effect of electron correlation on optical absorption spectra}
\label{sec:effect_of_correlation}
\begin{figure}[h]
\centering
\includegraphics[width=0.3\textwidth]{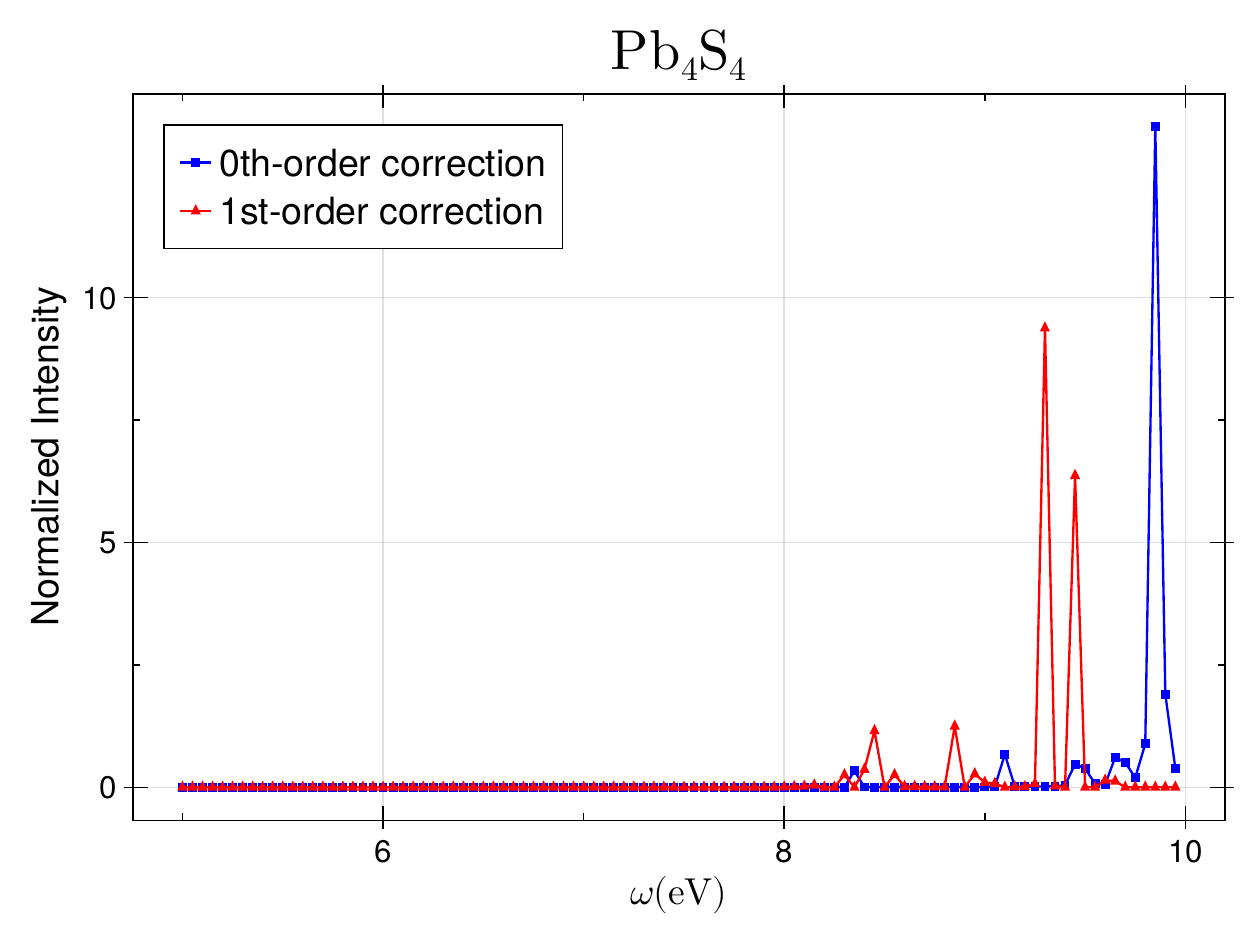}
\includegraphics[width=0.3\textwidth]{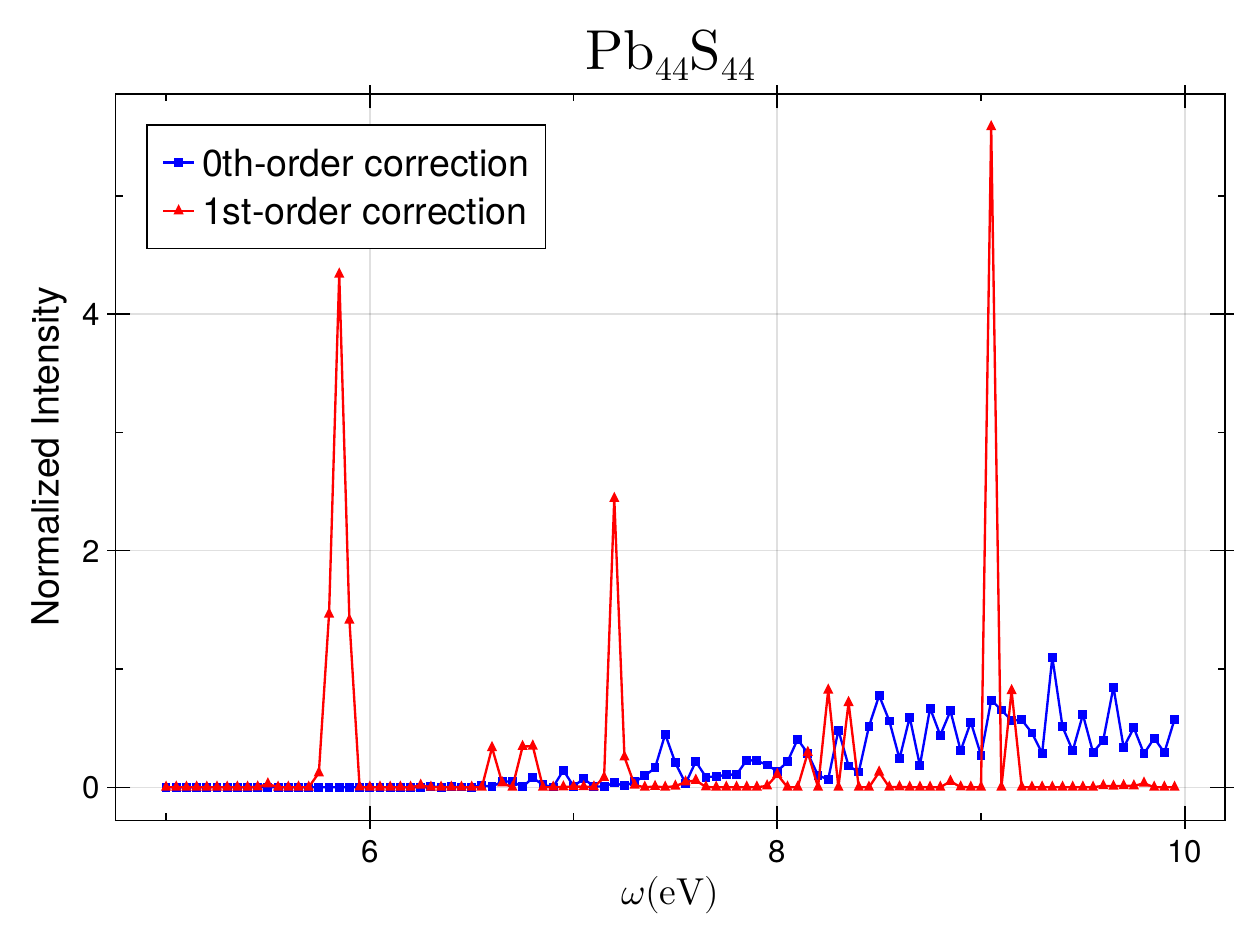}
\includegraphics[width=0.3\textwidth]{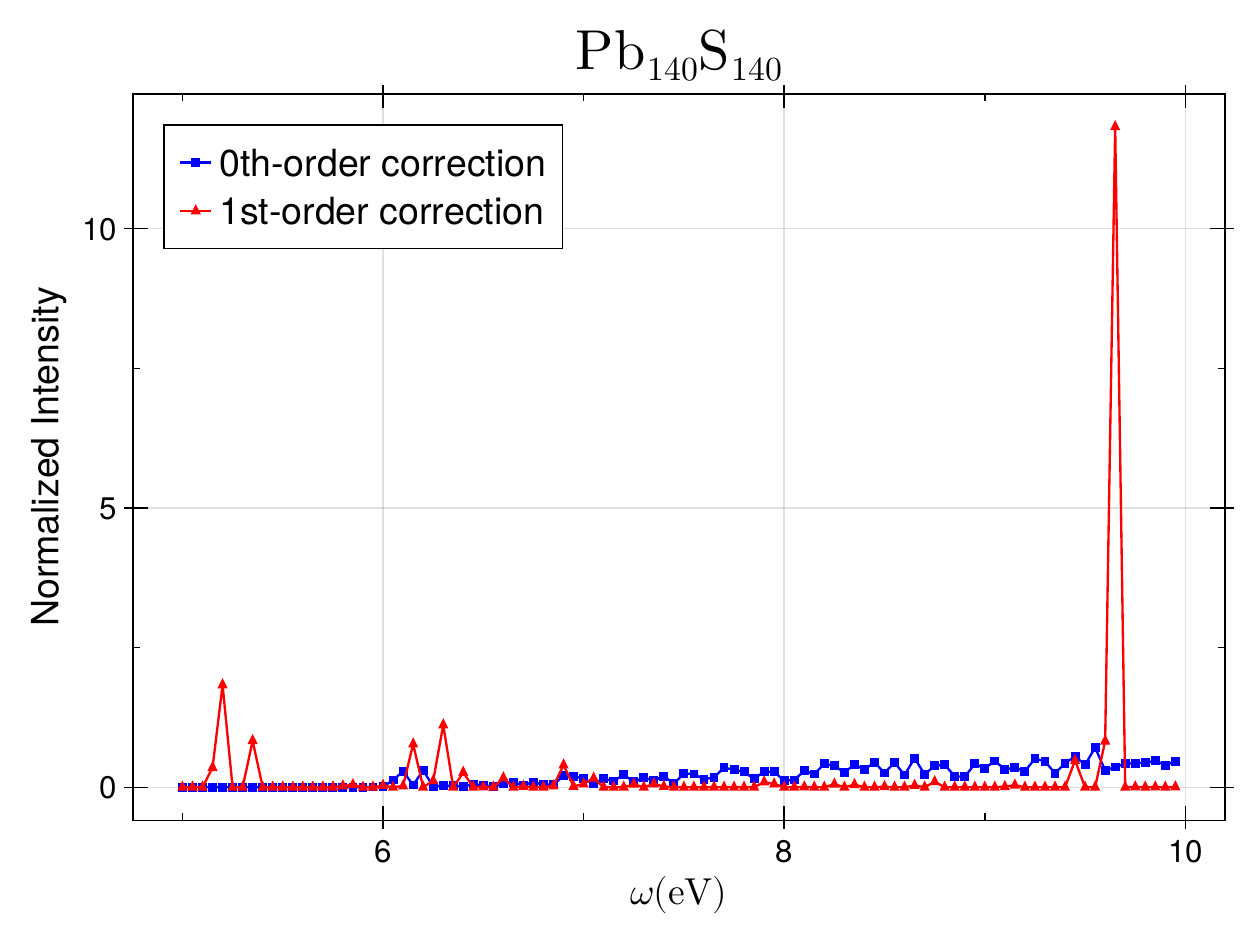}
\caption{\centering Absorption spectra of the PbS QDs from the polarization propagator: Hartree–Fock (HF, blue) and first-order corrected (red). First-order terms \(A^{(1)}\) and \(B^{(1)}\) produce a pronounced red-shift of the dominant transitions.}
\label{fig:pbs_1st_comb}
\end{figure}

\begin{figure}[h]
\centering
\includegraphics[width=0.45\textwidth]{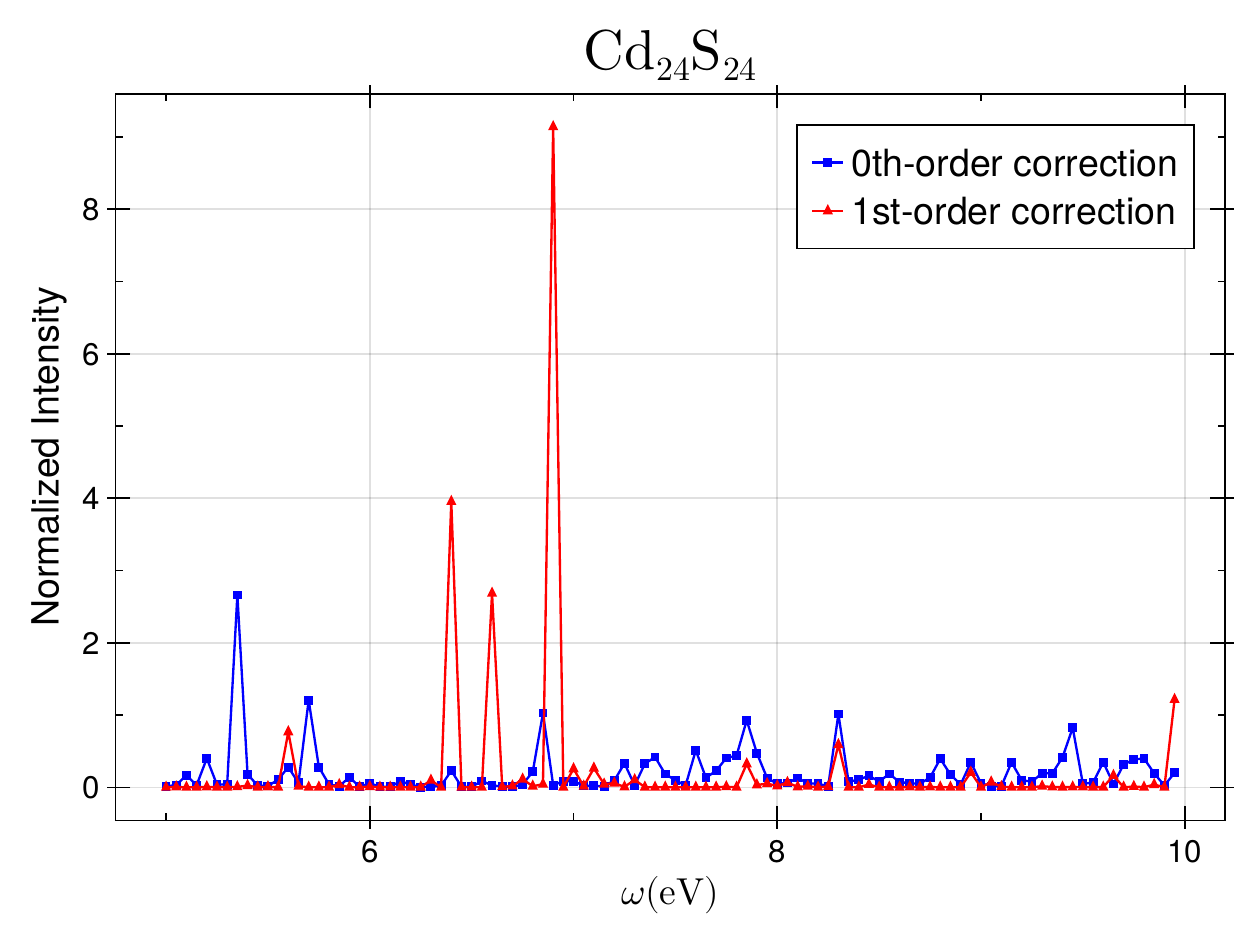}
\includegraphics[width=0.45\textwidth]{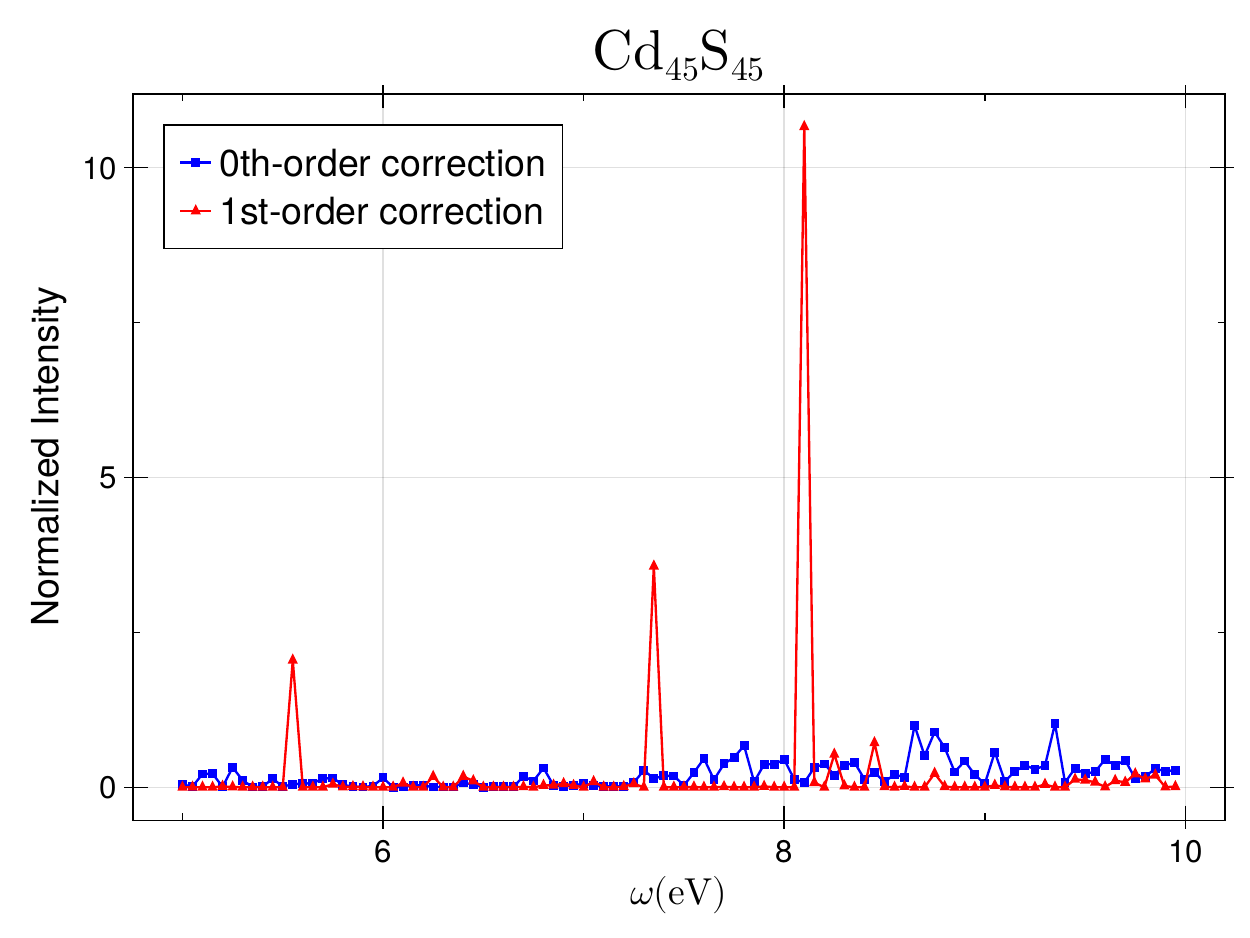}
\caption{\centering Absorption spectra of the CdS QDs from the polarization propagator: Hartree–Fock (HF, blue) and first-order corrected (red). First-order terms \(A^{(1)}\) and \(B^{(1)}\) produce a pronounced red-shift of the dominant transitions.}
\label{fig:cds_1st_comb}
\end{figure}

The effects of electron correlation have been shown to dramatically change calculated properties of materials\cite{Guo2022correlated,Chernov2023correlated}. Including first order correlated effects in this context means calculating the M-operator elements at first order, $\mathbf{A}^{(1)},\mathbf{B}^{(1)}$, and $\mathbf{S}^{(1)}$. Doing the first order correction involves calculation of six-dimensional spatial integrals from \autoref{eq:a(1)} and \autoref{eq:b(1)}, which are solvable using the method presented by Boys et al\cite{boys1950}. This correction introduces correlation into the results, and has a distinct effect on the optical properties. \autoref{fig:pbs_1st_comb} and \autoref{fig:cds_1st_comb} show the first order corrected spectra and HF spectra overlaid, for a visual representation of what \autoref{tab:max_peaks_pbs} and \autoref{tab:max_peaks_cds} shows quantitatively.
Each chemical system behaves differently in regards to the addition of correlated effects, but across all systems it is true that including correlation at any level results in significant changes to the optical spectra compared to the HF spectra. One result seen across systems is the shift from a low probability, distributed spectra to a few high probability transitions. In $\ce{Cd_{24}S_{24}}$ the HF spectra shows little transition probability in the scanned range of frequencies, while the correlated spectra has three dominant transitions in the same range. $\ce{Cd_{45}S_{45}}$ shows similar results, though the location of the dominant transitions shift relative to the smaller system. $\ce{Pb_4S4}$ shows different results from the others, due to having a strong high energy transition in the uncorrelated spectra. The correlated results here show that the dominant transition is only 70\% the strength of the uncorrelated dominant peak, a result that is unique to this system. $\ce{Pb_{44}S_{44}}$ and $\ce{Pb_{140}S_{140}}$ each show consistent results with the CdS systems, where the uncorrelated spectra has much lower intensity than the correlated counterpart. A consistent result across all PbS systems is the dominant transition in the high energy regime, while other peaks appear at various values of $\omega$ the high energy (9+ eV) regime is where the most probable transition is located for all system sizes.
\newpage
\subsection{Higher order correlated effects on optical absorption spectra}
\label{sec:ho_effects}
\begin{figure}[h]
\centering
\includegraphics[width=0.3\textwidth]{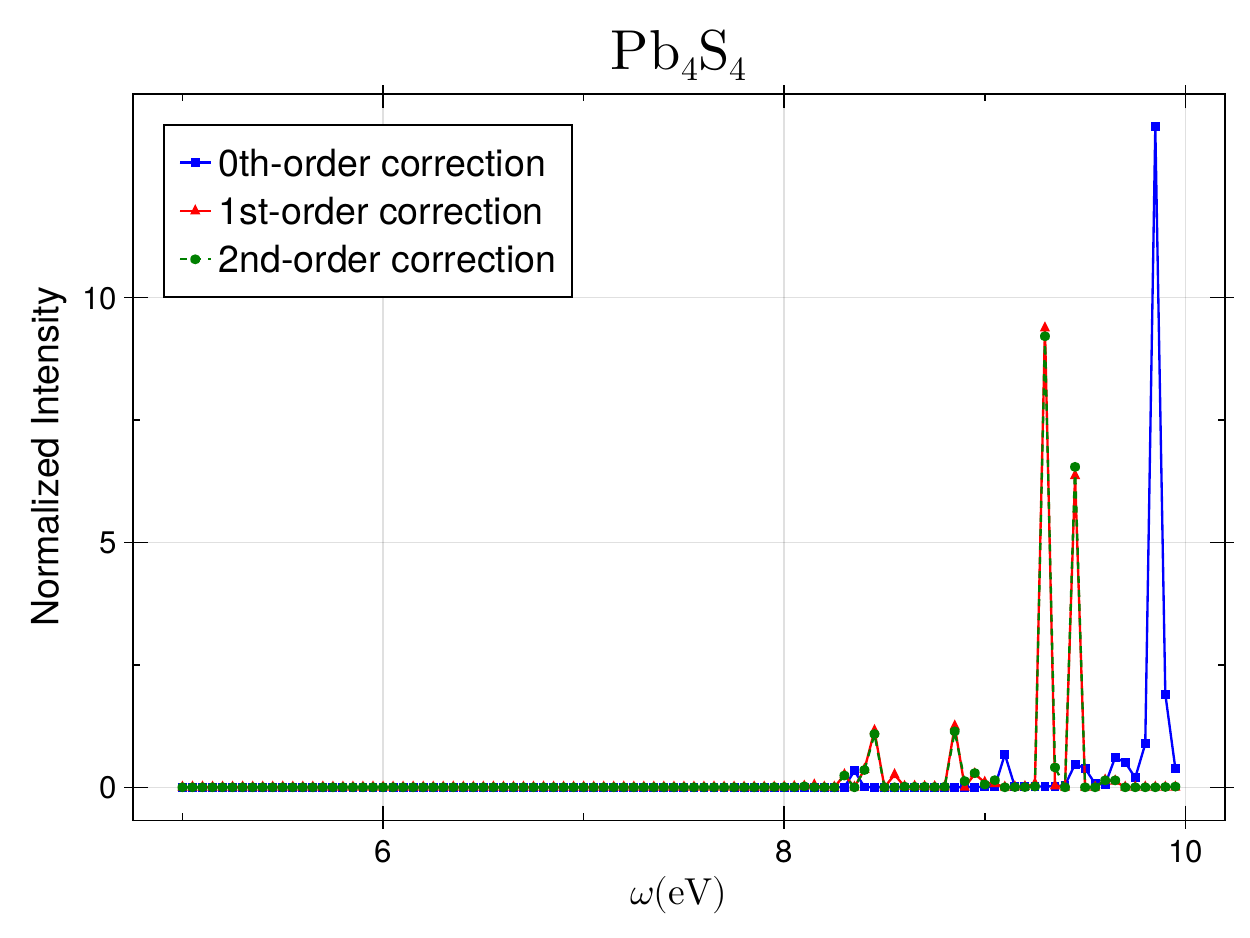}
\includegraphics[width=0.3\textwidth]{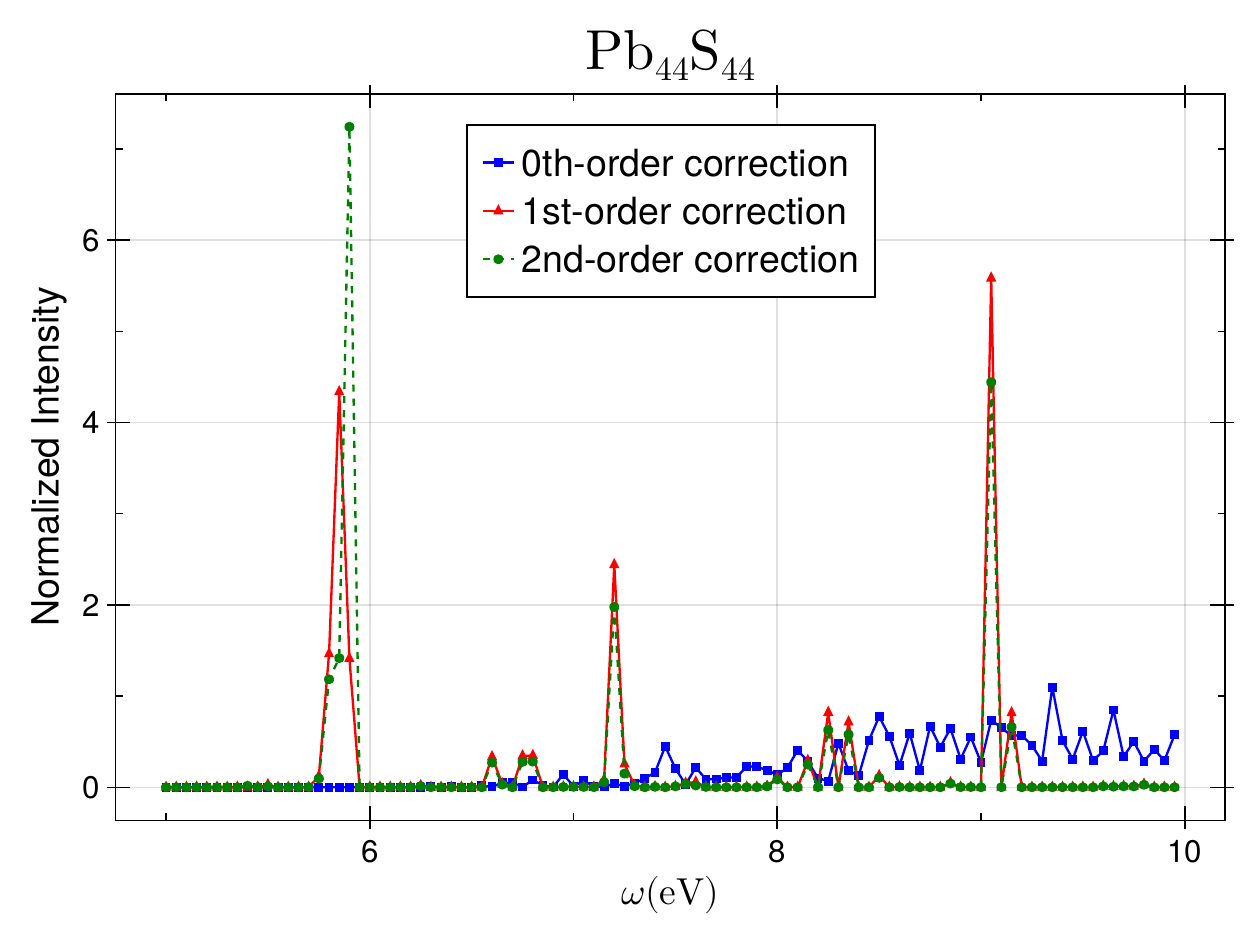}
\includegraphics[width=0.3\textwidth]{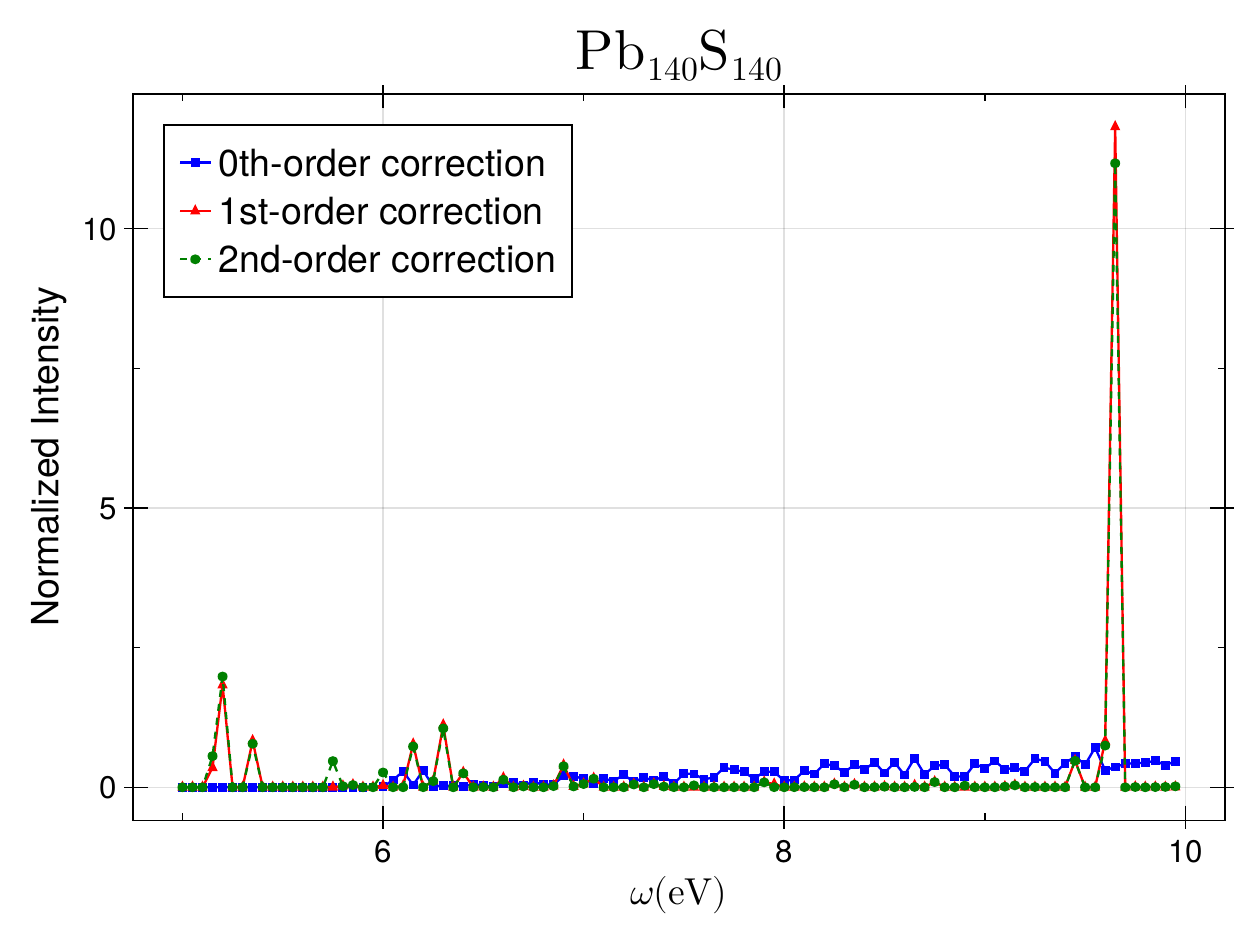}
\caption{\centering Pbs absorption spectra comparing HF (blue), first-order (red), and second-order (green) results. Second-order corrections \((S^{(2)},A^{(2)},B^{(2)})\) yield small additional energy shifts relative to first order but introduce new low probability transitions as well as change the relative intensities of the dominant transitions.}
\label{fig:pbs_2nd_comb}
\end{figure}

\begin{figure}[h]
\centering
\includegraphics[width=0.45\textwidth]{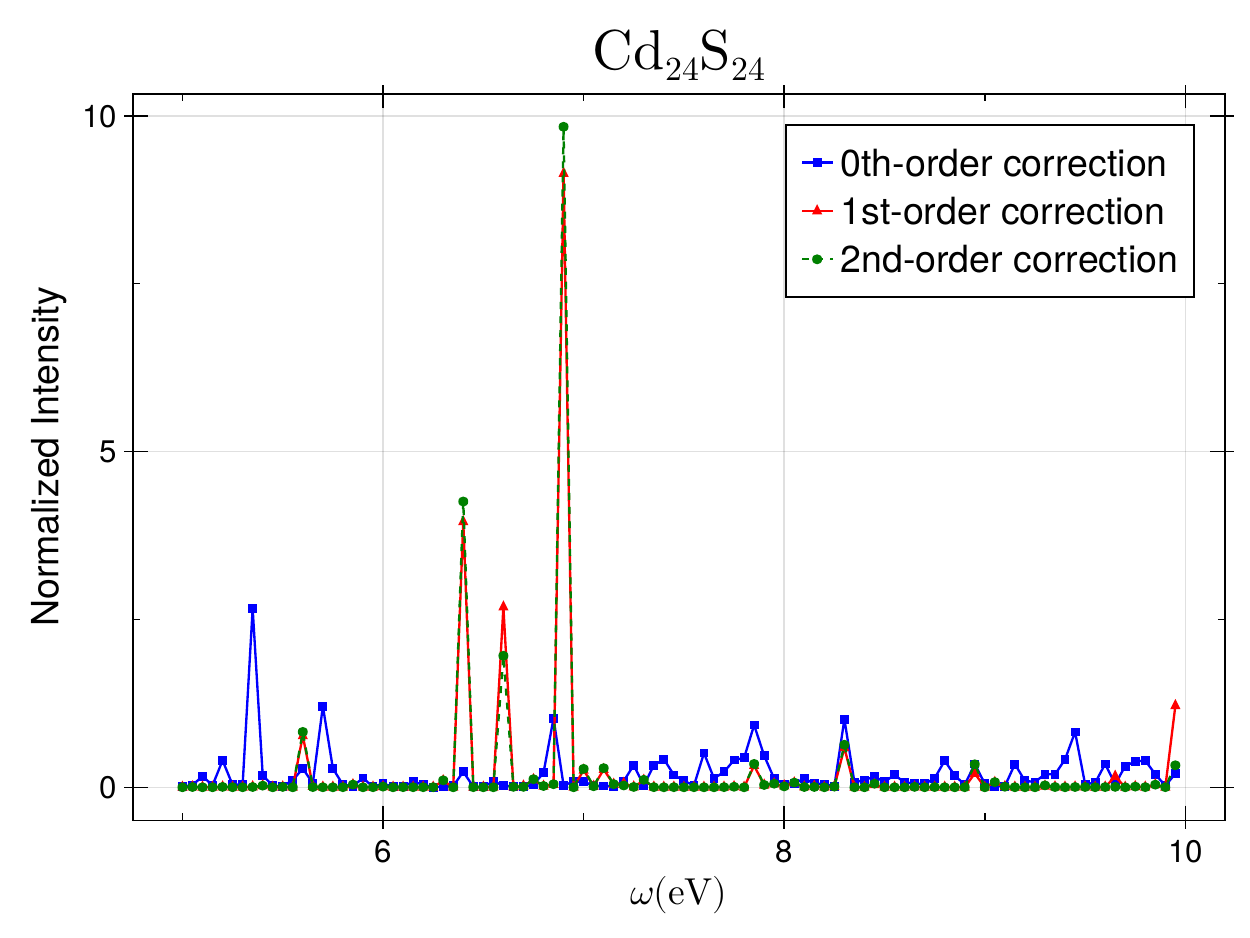}
\includegraphics[width=0.45\textwidth]{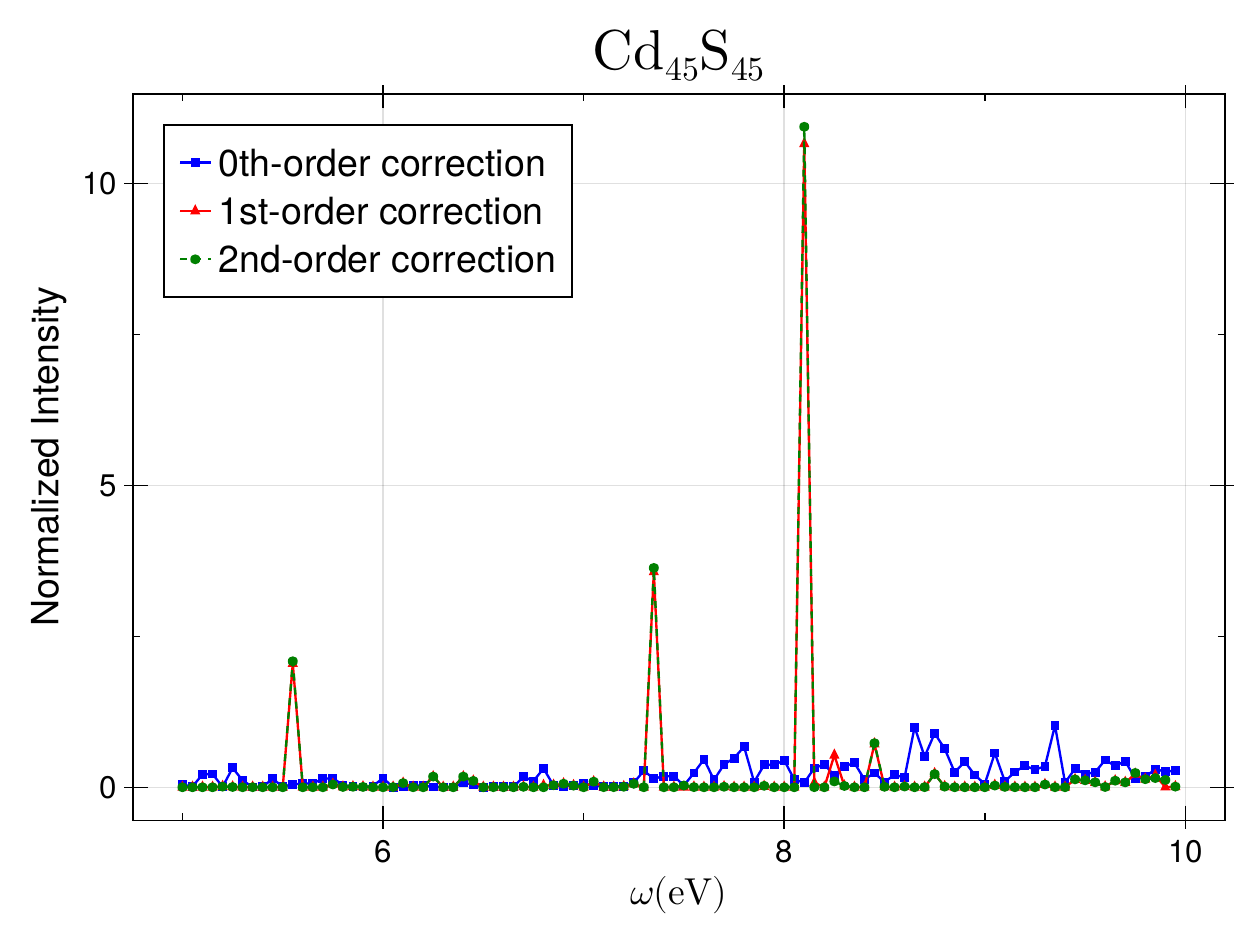}
\caption{\centering CdS absorption spectra comparing HF (blue), first-order (red), and second-order (green) results. Second-order corrections \((S^{(2)},A^{(2)},B^{(2)})\) yield small additional energy shifts relative to first order but introduce new low probability transitions as well as change the relative intensities of the dominant transitions.}
\label{fig:cds_2nd_comb}
\end{figure}

The effects of first order correlation in \autoref{sec:effect_of_correlation} cannot be understated, and they represent a dramatic deviation from the uncorrelated results. This makes including first order correlation necessary for accuracy.Going to higher order corrections in correlation requires increasingly expensive computational resources, and as shown in \autoref{tab:amat_size} the size of the space to sample scales from $\sim10^4$ in first order to $\sim10^{10}$ in second order. In \autoref{sec:clock_time} we show the relative computational time needed for calculation of the $2^{\mathrm{nd}}$ order correction to the PP and how our method compares to full matrix diagonalization. Results are presented relative to the computational effort required for calculation of the $\mathrm{Pb}_4\mathrm{S}_4$. The increasing computational complexity highlights the need for proper diagrammatic evaluation of the matrix elements of $\mathbf{A},\mathbf{B}$ and $\mathbf{S}$. There are many elements of the second order corrections given in \autoref{sec:pp_def}, but only the terms shown in \autoref{sec:diagrams} contribute to the results. This truncation makes the contributions from second order corrections possible to calculate, though the integrals are still computationally challenging. Using the procedure described in \autoref{sec:krylov_theory} we are able to generate the $2^{\mathrm{nd}}$ order corrected results. In \autoref{fig:pbs_2nd_comb} and \autoref{fig:cds_2nd_comb} we show the results as an overlay on the uncorrelated spectra, and first order corrected spectra.
In general, the second order terms provide small corrections to the first order results. The exception to this is in the $\ce{Pb_{44}S_{44}}$ system where the dominant transition changes from a high energy transition to low energy, indicating that second order results play a more important role in the energy landscape of this system. The results from the second order calculation can be seen in \autoref{tab:max_peaks_pbs} and \autoref{tab:max_peaks_cds}, both of which include the calculation wall clock time. Despite the efficiency gains from diagrammatics, the computational cost of these results is still high and require upwards of $\approx1$ day for the spectra to be calculated, so more efficient methods are needed. Prioritizing accuracy in probing the energy landscapes of these structures requires going to higher orders of correction, and these results demonstrate the efficacy of diagrammatic evaluation in conjunction with the folded spectrum method in polarization propagator calculations.
\section{Discussion}
\subsection{Analysis of the excitonic density of states}
\label{sec:density_of_states}
\begin{figure}[h!]
\centering
\includegraphics[scale=0.7]{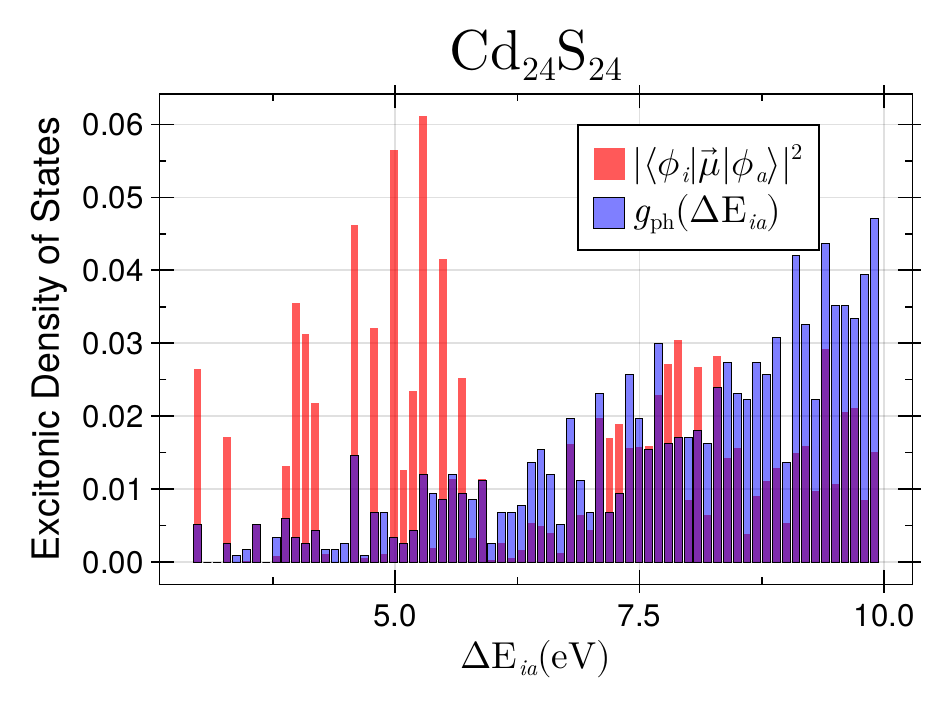}
\caption{\centering Excitonic density of states histogram overlaid with the square of the dipole operator for the Hartree Fock reference orbitals of the $\mathrm{Cd}_{24}\mathrm{S}_{24}$ quantum dot. Energy spacing $\Delta E_{ia}$ is $0.1$ eV for each bin.}
\label{cd24_dos}
\end{figure}

\begin{figure}[h!]
\centering
\includegraphics[scale=0.7]{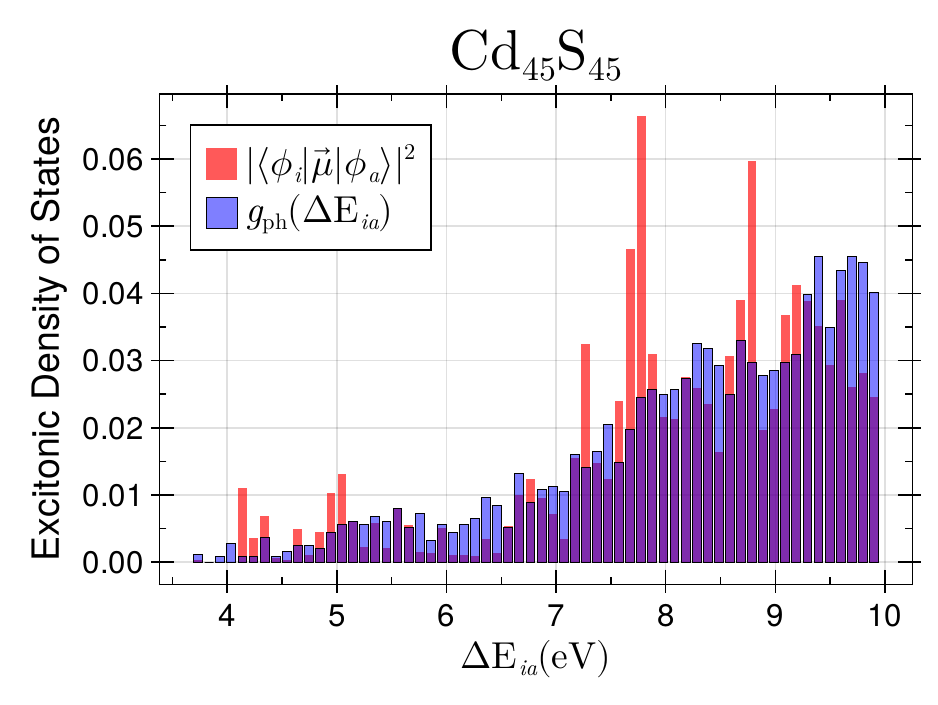}
\caption{\centering Excitonic density of states histogram overlaid with the square of the dipole operator for the Hartree Fock reference orbitals of the $\mathrm{Cd}_{45}\mathrm{S}_{45}$ quantum dot. Energy spacing $\Delta E_{ia}$ is $0.1$ eV for each bin.}
\label{cd45_dos}
\end{figure}

\begin{figure}[h!]
\centering
\includegraphics[scale=0.7]{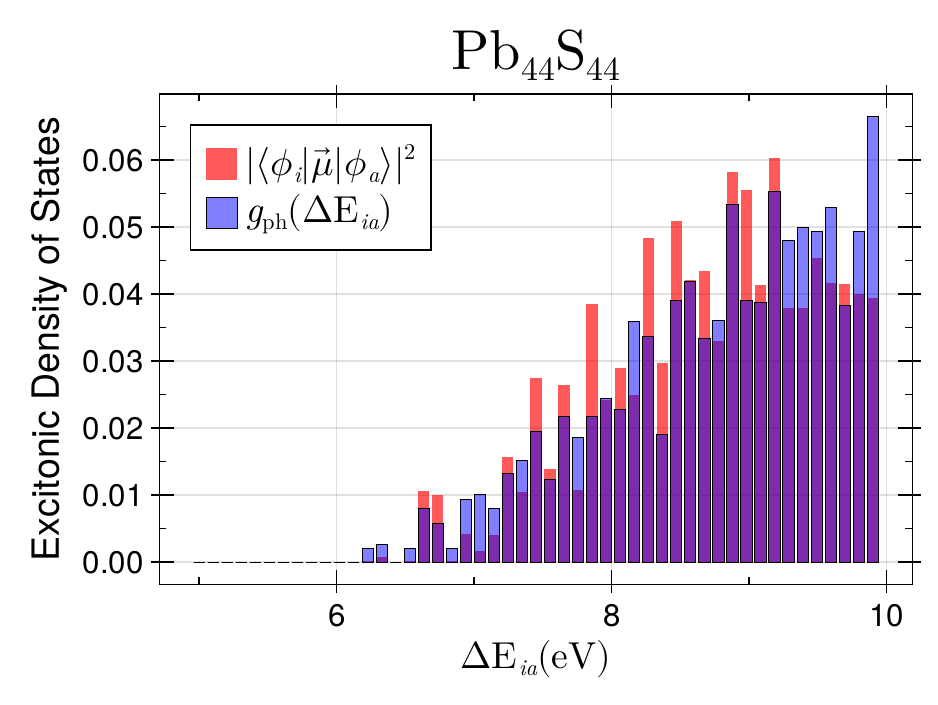}
\caption{\centering Excitonic density of states histogram overlaid with the square of the dipole operator for the Hartree Fock reference orbitals for the $\mathrm{Pb}_{44}\mathrm{S}_{44}$ quantum dot. Energy spacing $\Delta E_{ia}$ is $0.1$ a.u. for each bin.}
\label{pb44_dos}
\end{figure}

\begin{figure}[h!]
\centering
\includegraphics[scale=0.7]{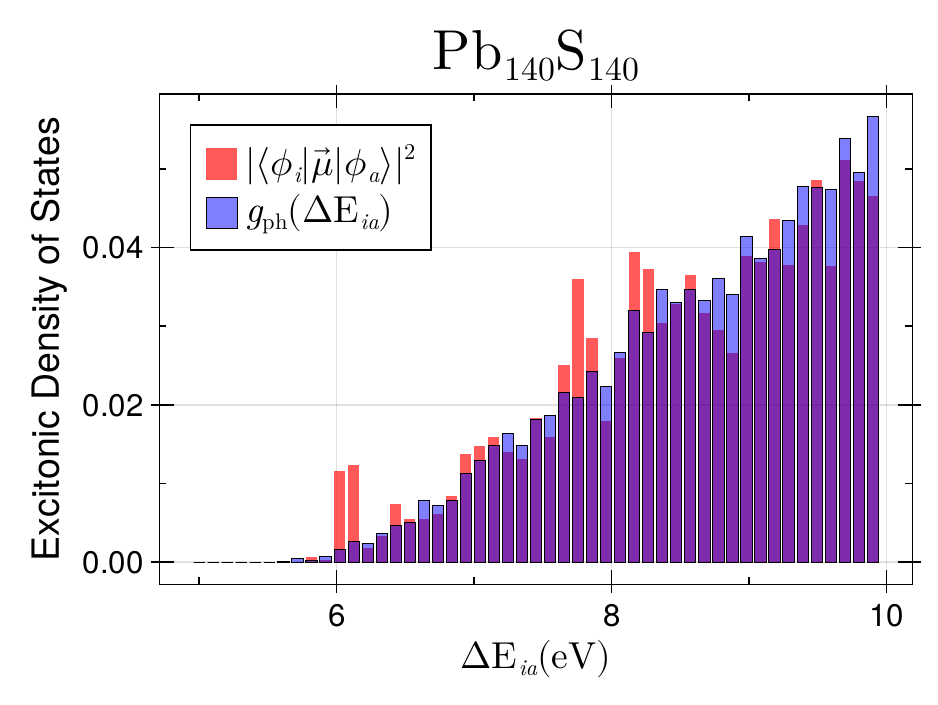}
\caption{\centering Excitonic density of states histogram overlaid with the square of the dipole operator for the Hartree Fock reference orbitals for the $\mathrm{Pb}_{140}\mathrm{S}_{140}$ quantum dot. Energy spacing $\Delta E_{ia}$ is $0.1$ a.u. for each bin.}
\label{pb140_dos}
\end{figure}

One of the important pieces to any excited state calculation is the density of states $g$. The density of states typically contains information about the energy states of a system, but we are primarily interested in the excitonic states which are defined by
\begin{align}
    E_a - E_i = \Delta E_{ia} .
\end{align}
Using the single particle states calculated from TERACHEM, we can find  each $\Delta E_{ia}$ and create an excitonic density of states $g_{\mathrm{ph}}(\Delta E_{ia})$ which counts the number of particle-hole states in a given $\Delta E$ range. The excitonic density of states shows the number of accessible transitions in a given energy range, but we are also interested in the probability of a given transition occurring. The probability of a transition taking place is given by the square of the dipole operator
\begin{align}
    \mu^2_{ia} = \vert \langle \phi_i \vert \vec{\mu} \vert \phi_a \rangle \vert^2
\end{align}
which shows how strongly coupled state $i$ and state $a$ are in the presence of an external electromagnetic field. This quantity, in combination with the excitonic density of states, shows which transitions are most likely to occur in a system. A large density of states for a given $\Delta E$ means more states are accessible for that incoming photon, but if those states have weak coupling with the electromagnetic field then a transition is still unlikely to occur.\autoref{cd24_dos} and \autoref{cd45_dos} show a histogram of both $g_{\mathrm{ph}}(\Delta E_{ia})$ and $\mu^2_{ia}$ for the CdS systems. 
\newpage
\autoref{pb44_dos} and \autoref{pb140_dos} show the same results for the PbS systems, though for the PbS systems specifically the range of $\Delta E_{ia}$ was restricted to be between $5-10$ eV due to zero possible excitations below the $5$ eV threshold. Note that the figure for the $\mathrm{Pb}_4\mathrm{S}_4$ system is not shown, this is due to a low number of excitonic states resulting in a sparse density of states. The regions of these plots where both $g_{\mathrm{ph}}$ and $\mu^2_{ia}$ are high are the energy values that contribute most to the spectra shown in \autoref{sec:results} because these energies have both strong coupling and a high number of possible excitations.
%
%\begin{figure}[h!]
%\centering
%\includegraphics[scale=0.7]%{figs/density_of_states/pb4_ediff_hist.pdf}
%\caption{\centering Excitonic density of states histogram overlaid with the square of the dipole operator for the Hartree Fock reference orbitals for the $\mathrm{Pb}_{4}\mathrm{S}_{4}$ quantum dot. Energy spacing $\Delta E_{ia}$ is $0.1$ a.u. for each bin.}
%\label{pb4_dos}
%\end{figure} 
%
These plots show the most important underlying features that go into generating the spectra from \autoref{sec:results}, and indicate exactly where computational effort should be spent. Future work could utilize this data to make stronger approximations to absorption spectra, by providing a ranking of the most important energy ranges to calculate.
\newpage
\subsection{Error analysis of the interior eigenvectors}
In order to test the effectiveness of the inverse Krylov subspace + folded spectrum method in isolating the interior eigenvectors
of the polarization propagator matrix, we compute the variance using eigenvectors obtained from the folded spectrum method. 
For any vector $\mathbf{x}$ and a matrix $\mathbf{A}$, the variance in terms of the Rayleigh coefficient is defined as
\begin{align}
    \sigma^2(\mathbf{x}) = \rho_{A^2}(\mathbf{x}) - \rho_{A}^2(\mathbf{x})
\end{align}
where $\rho_{A}(\mathbf{x}) = \langle \mathbf{x} \vert A \vert \mathbf{x} \rangle$ and $\rho_{A^2}(\mathbf{x}) = \langle \mathbf{x} \vert A^2 \vert \mathbf{x} \rangle$, respectively. 
For an exact eigenvector $\mathbf{v}_k$ with eigenvalue $\lambda_k$, the variance $\sigma^2(\mathbf{v}_k) = 0$.
\par
\begin{figure}[h]
\centering
\includegraphics[scale=0.5]{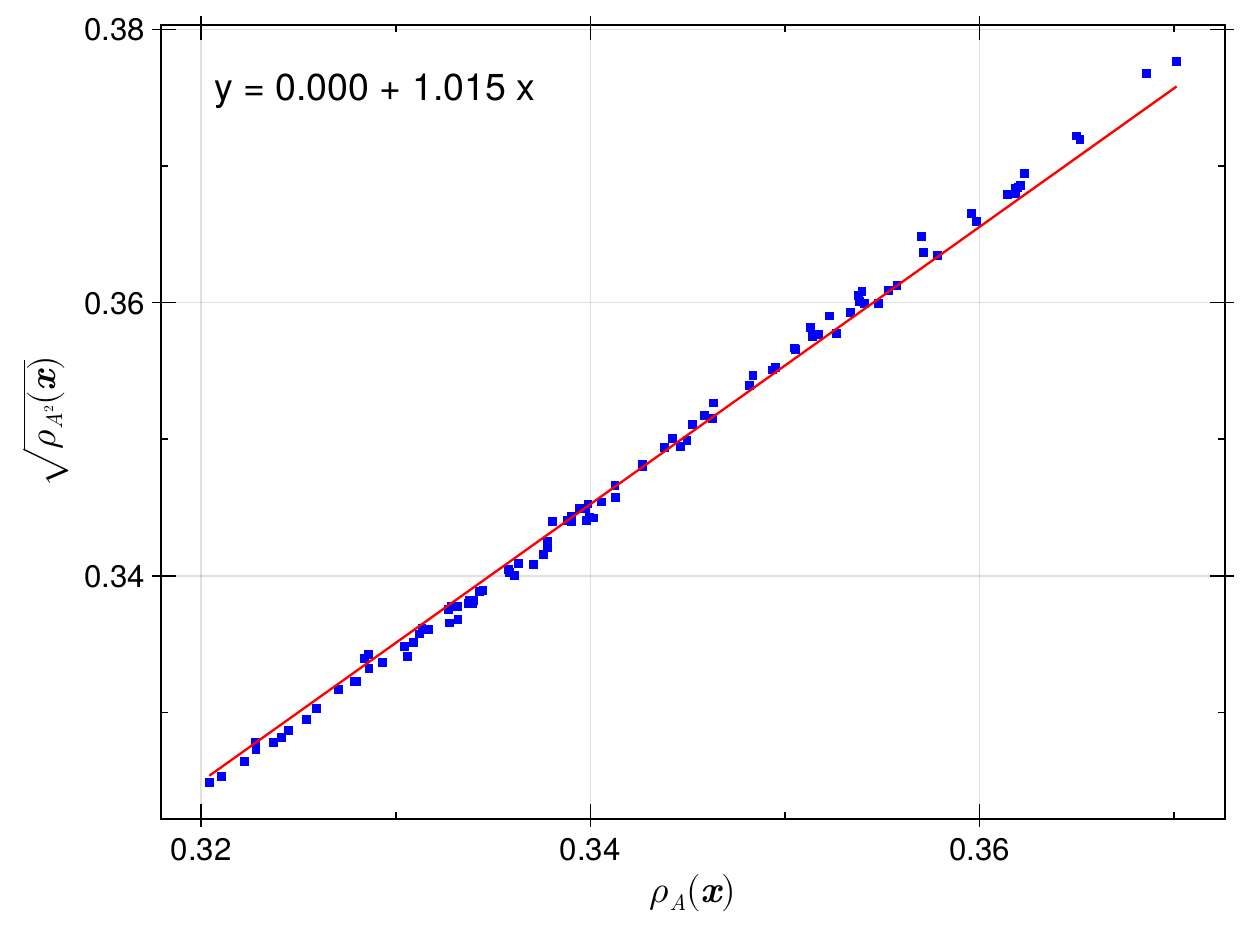}
\caption{\centering $\sqrt{\rho_{A^2}(\mathbf{x})}$  versus $\rho_{A}(\mathbf{x})$for the $\mathbf{A}$ matrix of $\mathrm{Pb}_4\mathrm{S_4}$. The regression line is shown in the upper left, with a slope of $m=1.015$ indicating that the folded spectrum method is constructing a good approximation of the exact eigenvectors.}
\label{fig:rho_comp}
\end{figure}

To assess the accuracy of the eigenvectors obtained using the procedure described in \autoref{sec:theory}, the linear relationship between $\rho_{A}(\mathbf{x})$ and $\sqrt{\rho_{A^2}(\mathbf{x})}$ was examined. This relationship is evaluated by plotting $\sqrt{\rho_{A^2}(\mathbf{x})}$ versus $\rho_{A}(\mathbf{x})$, as shown in \autoref{fig:rho_comp} for the $\mathbf{A}$ matrix constructed for the $\mathrm{Pb}_4\mathrm{S}_4$ system.
 As shown in \autoref{fig:rho_comp}, the slope of $m=1.015$ indicates that the folded spectrum is doing a good job approximating the exact eigenvectors $\mathbf{v}_k$ of the $\mathbf{A}$ matrix. This verifies that the folded spectrum method is appropriate to use for the calculations discussed in \autoref{sec:diagrams}, making it an efficient an accurate method for eigenvalue calculation of large matrices.
 \newpage
\subsection{Computational clock time comparison}
\label{sec:clock_time}
\begin{figure}[h]
\centering
\includegraphics[width=0.7\textwidth]{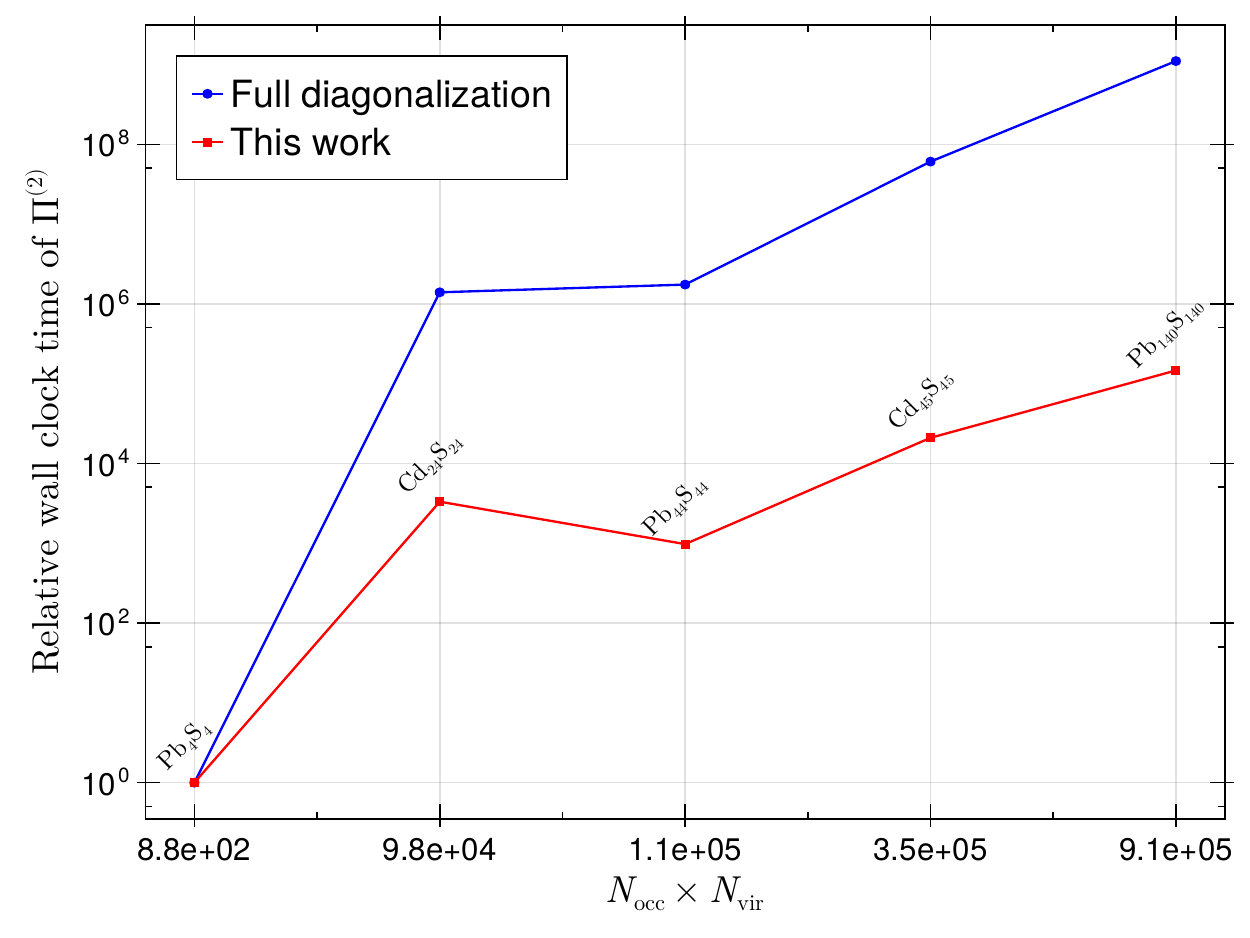}
\caption{\centering Relative computational effort required for all systems.}
\label{fig:rel_clock}
\end{figure}

In order to show that the methods developed in this work are effective, it is important to show that they result in a saving of real-world computational time. 
This is illustrated in \autoref{fig:rel_clock}, which shows the relative computational clock time for the present method as compared to the full diagonalization of the $\mathbf{M}$ matrix, whose cost scales as $\mathcal{O}(N^3)$ for an $N \times N$ matrix.  All values in this plot were normalized relative to the computational effort required to calculate results for $\mathrm{Pb}_4\mathrm{S}_4$, the smallest structure studied which has a computational time of $0.0013$ hours using our method. For example, the results in \autoref{fig:rel_clock} indicate that a standard matrix diagonalization procedure would require approximately $10^9$ times more computational effort for $\mathrm{Pb}_{140}\mathrm{S}_{140}$ than for $\mathrm{Pb}_4\mathrm{S}_4$. In contrast, the method developed in this work scales much more favorably, requiring only about $10^5$ times more computational effort over the same range of system sizes, resulting in a four-order-of-magnitude reduction in computational cost for the largest system.
 Also important to note in \autoref{fig:rel_clock} is that the full diagonalization is only including the computational effort required to diagonalize $\mathbf{M}$, and contains no information about the difficulty in constructing $\mathbf{M}$. Each matrix element in $\mathbf{M}$ is shown in \autoref{sec:diagrams} and requires the calculation of $S_{ia,jb}^{(2)}, A_{ia,jb}^{(2)}$, and $B_{ia,jb}^{(2)}$ for all $ia,jb$ terms, a total of $(N_{\mathrm{occ}}\times N_{\mathrm{vir}})^2$ terms. Therefore, the actual computational effort required using the conventional approach is much greater than indicated in \autoref{fig:rel_clock}. The computational effort in this work however does take into account the construction of the $\mathbf{M}$ matrix, using diagrammatic evaluation of the matrix elements as well as the inverse Krylov basis and folded spectrum method. Still we show significant real-time computational savings over just the matrix diagonalization procedure. Also notable is in \autoref{fig:rel_clock} the $\mathrm{Pb}_{44}\mathrm{S}_{44}$ system shows faster results than the $\mathrm{Cd}_{24}\mathrm{S}_{24}$ system, despite a comparable size of the excitation space. This is due to the structure of the $2^{\mathrm{nd}}$ order correction terms to the $\mathbf{M}$ matrix resulting in easier evaluation of the matrix elements. These results show that the method presented in this work offers significant real-time computational savings, making it effective for calculation of high order correction terms in the polarization propagator.
 \newpage
\subsection{Connection to the Riesz projector}
\label{sec:riesz_proj}
Formally, the Riesz projector $\mathbf{P}_\Gamma$ associated with a matrix $\mathbf{M}$ is defined as
\begin{align}
\label{eq:riesz_projector}
\mathbf{P}_\Gamma = \frac{1}{2\pi i} \oint_{\Gamma} (z\mathbf{I}-\mathbf{M})^{-1} dz,
\end{align}
where $\Gamma$ is a closed contour in the complex plane that encloses the eigenvalues of interest. Applying $\mathbf{P}_\Gamma$ to a vector $v$ extracts the component of $v$ in the subspace of the enclosed eigenvalue. If an eigenvalue $\lambda$ is the only eigenvalue of $\mathbf{M}$ inside the region $\Gamma$ then the Riesz projector instead is denoted as $\mathbf{P}_\lambda$. The Riesz projector offers much of the same benefits as the inverted Krylov subspace, by defining a subspace nearest to the eigenvalue of interest, and rotating the basis as a function of $\omega$. However, the contour integration lacks the linear algebraic potential for approximations, such as those discussed in \autoref{sec:krylov_theory}. By applying simple linear algebraic transformations such as expressing a matrix $\mathbf{M}$ by its diagonal and off-diagonal components $\mathbf{D}$ and $\mathbf{F}$ we can create a computationally efficient algorithm. Both the Riesz projector and folded spectrum method solve the interior eigenvalue problem however, and either could have been used for the purpose of solving the PP.
\section*{Conclusion}
In this work we have presented a computationally efficient framework for calculation of the frequency dependent polarization propagator in semiconductor quantum dots using an inverted Krylov subspace based approach. This approach enables selective calculation of the eigenvalues of the resolvent Hamiltonian superoperator, leading to excitation energies without needing full diagonalization of the many-body Hamiltonian. Through perturbative expansion up to $2^{\mathrm{nd}}$ order we have captured essential electron correlation information, and through identification and elimination of non-contributing terms in the diagrammatic perturbative expansion we have made calculation of these $2^{\mathrm{nd}}$ order terms computationally tractable. Using these techniques, we are able to show that $1^{\mathrm{st}}$ order corrections generally red shift spectral peaks, and enhance emission intensities, while $2^{\mathrm{nd}}$ order corrections offer small refinements on the absorption spectra of these materials. Our calculations also reveal various size-dependent quantities behavior in both lead selenide (PbS) and cadmium selenide (CdS) quantum dots. CdS quantum dots exhibit a blue-shifting effect on absorption peaks with increasing diameter, while PbS quantum dots show strong transitions at high energies at both large and small diameters, but a spreading of the dominant transitions in the intermediate diameter regime. Additionally, the computational scaling shown in \autoref{sec:clock_time} indicates the strength of the inverse Krylov subspace approach to calculation of these excited state energies using the folded spectrum method. The range of calculation time from order seconds to $\approx 190$ hours for $\mathrm{Pb}_{140}\mathrm{S}_{140}$ shows the challenges of many-body computations on these large quantum dots, though still offers significant performance improvement over explicit construction and diagonalization of the polarization propagator.
%\par
%%%%%%%%%%%%%%%%%%%% Future outlook? %%%%%%%%%%%%%%%%%%%%%
% While our $2^{\mathrm{nd}}$ order results capture the essential physics from these direct benchmarking against higher order methods such as ADC(3), or EOM-CCSD for these systems is a priority for future work.
%\section*{Appendix}
%\input{sections/008_appendix}
\section*{Acknowledgments}
This research was supported by the National Science Foundation under Grant No. CHE-2102437, CHE-1950802, ACI-1341006, ACI-1541396 and by computational resources provided by Syracuse University. 
\clearpage
\newpage
\newpage
\bibliography{mybib}
\end{document}